\documentclass[prx,aps,twocolumn,longbibliography]{revtex4-2}

\usepackage[paperwidth=210mm,paperheight=297mm,centering,hmargin=1.65cm,vmargin=2.04cm]{geometry}

\pdfoutput=1

\newcommand{\trf}{t_{\textrm{rf}}}
\newcommand{\kB}{k_{\textrm{B}}}

\newcommand{\UD}{U_{\textrm{D}}}

\newcommand{\Bres}{B_{\textrm{res}}}
\newcommand{\Ep}{E_{\textrm{p}}}
\newcommand{\Ed}{E_{\textrm{d}}}
\newcommand{\ab}{a_{\textrm{b}}}
\newcommand{\ai}{a_{\textrm{i}}}

\usepackage{float}
\usepackage{fourier}
\usepackage{amsmath,amssymb}
\usepackage{mathtools}
\usepackage{amsthm}
\usepackage{empheq}
\usepackage{cases}
\usepackage{times}
\usepackage{upgreek}
\usepackage{psfrag} 
\usepackage{latexsym} 
\usepackage{amstext}
\usepackage{amsfonts}
\usepackage{amsxtra} 
\usepackage{dcolumn}
\usepackage{dcolumn}
\usepackage{textcomp}
\usepackage{graphicx}
\usepackage{bm}
\usepackage{braket}
\usepackage{units}
\usepackage[svgnames,dvipsnames]{xcolor}

\definecolor{myColor}{rgb}{0.02,0.12,0.3}
\definecolor{myciteColor}{rgb}{0.39,0.7,0.89}
\usepackage[colorlinks=true,citecolor=myColor,linkcolor=myColor,urlcolor=myColor]{hyperref}

\def\be{\begin{equation}}
\def\ee{\end{equation}}

\def\nobreakbefore{%
  \relax\ifvmode\else
    \ifhmode
      \ifdim\lastskip > 0pt\relax
        \unskip\nobreakspace
      \else 
        \nobreakspace
      \fi
    \fi
  \fi
}
\let\oldcite\cite
\renewcommand\cite{\nobreakbefore\oldcite}

\makeatletter
\def\@fnsymbol#1{\ensuremath{\ifcase#1\or *\or \dagger\or \ddagger\or
   \mathsection\or \mathparagraph\or \|\or **\or \dagger\dagger
   \or \ddagger\ddagger \else\@ctrerr\fi}}
\makeatother

\begin{document} 
 
\title{
Universal quantum dynamics of Bose polarons\\
}
\author{
Ji\v{r}\'{i}~Etrych,$^{{\color{myColor}}}$
Gevorg Martirosyan,$^{{\color{myColor}}}$
Alec Cao,$^{{\color{myColor}\dag}}$
Christopher J. Ho,$^{}$
Zoran Hadzibabic,$^{}$
and Christoph~Eigen$^{\ast}$
}
\affiliation{
Cavendish Laboratory, University of Cambridge, J. J. Thomson Avenue, Cambridge CB3 0HE, United Kingdom}

\begin{abstract}
Predicting the emergent properties of impurities immersed in a quantum bath is a fundamental challenge that can defy quasiparticle treatments.
Here, we measure the spectral properties and real-time dynamics of mobile impurities injected into a weakly interacting homogeneous Bose--Einstein condensate, using two broad Feshbach resonances to tune both the impurity-bath and intrabath interactions. 
For attractive impurity-bath interactions, the impurity spectrum features a single branch, which away from the resonance corresponds to a well-defined attractive polaron; near the resonance we observe dramatic broadening of this branch, suggesting a breakdown of the quasiparticle picture. 
For repulsive impurity-bath interactions, the spectrum features two branches: the attractive branch that is dominated by excitations with energy close to that of the Feshbach dimer, but has a many-body character, and the repulsive polaron branch.
Our measurements show that the behavior of impurities in weakly interacting baths is remarkably universal, controlled only by the bath density and a single dimensionless interaction parameter.
\end{abstract}
\maketitle 

\section{Introduction}
\vspace{-1em}
Understanding strongly correlated quantum matter is a fundamental goal of many-body physics.
Remarkably, complex systems with many interacting degrees of freedom, from Fermi liquids\cite{Baym:2004c} to superfluids~\cite{Pitaevskii:2016}, are often amenable to relatively simple quasiparticle descriptions.
The polaron, a mobile impurity dressed by the excitations of a medium, is a paradigmatic quasiparticle, originally conceived to describe electrons moving through a crystal~\cite{Landau:1933,Pekar:1946a} and now relevant in many contexts, from condensed matter~\cite{Alexandrov:2008} to surface chemistry~\cite{Franchini:2021} and quantum computation~\cite{Tomza:2019}.

Experiments with neutral ultracold atoms have served as a powerful platform for studying Fermi~\cite{Schirotzek:2009,Nascimbene:2009,Kohstall:2012,Koschorreck:2012,Zhang:2012,Wenz:2013,Cetina:2016,Scazza:2017,Darkwah:2019,Yan:2019,Ness:2020,Baroni:2024,Vivanco:2025} and Bose~\cite{Hu:2016,Jorgensen:2016,Rentrop:2016,Yan:2020,Skou:2021,Skou:2022,Cayla:2023,Morgen:2025,semiconpolaron} polarons, where impurity atoms are coupled to a spin-polarized Fermi sea or a Bose--Einstein condensate (BEC). Crucially, impurity-bath interactions (characterized by the scattering length $a$) can be tuned to be either attractive or repulsive, and either weak or strong, by exploiting Feshbach resonances~\cite{Chin:2010} associated with a bound state of the impurity and a bath atom. 
In vacuum, this dimer state (with reduced mass $m_r$) has energy $\Ed=-\hbar^2/(2 m_r a^2)$ for $a>0$ and becomes unbound on resonance ($a\to \infty$), while in a medium it may instead connect to the negative-energy attractive polaron at $a<0$~\cite{Rath:2013,Christianen:2024}.
For the Fermi polaron, it is established that the quasiparticle picture holds even in the strongly interacting regime~\cite{Massignan:2014,Parish:2023}, while in the Bose case, where the impurity can dramatically distort the more compressible bath, the validity of a quasiparticle description is an open question~\cite{Shchadilova:2016,Grusdt:2017,Guenther:2021,Christianen:2022}.
Moreover, while a spin-polarized atomic Fermi sea is typically noninteracting, in the case of a BEC, both repulsive intrabath interactions (characterized by the scattering length $\ab$) and three-body Efimov correlations make the problem more intricate~\cite{Levinsen:2015,Ardila:2015,Shchadilova:2016,Grusdt:2017,Sun:2017,Yoshida:2018,Drescher:2020,Guenther:2021,Massignan:2021,Christianen:2022,Schmidt:2022,Christianen:2024,Mostaan:2023}.

\begin{figure}[t!]
\centerline{\includegraphics[width=1\columnwidth]{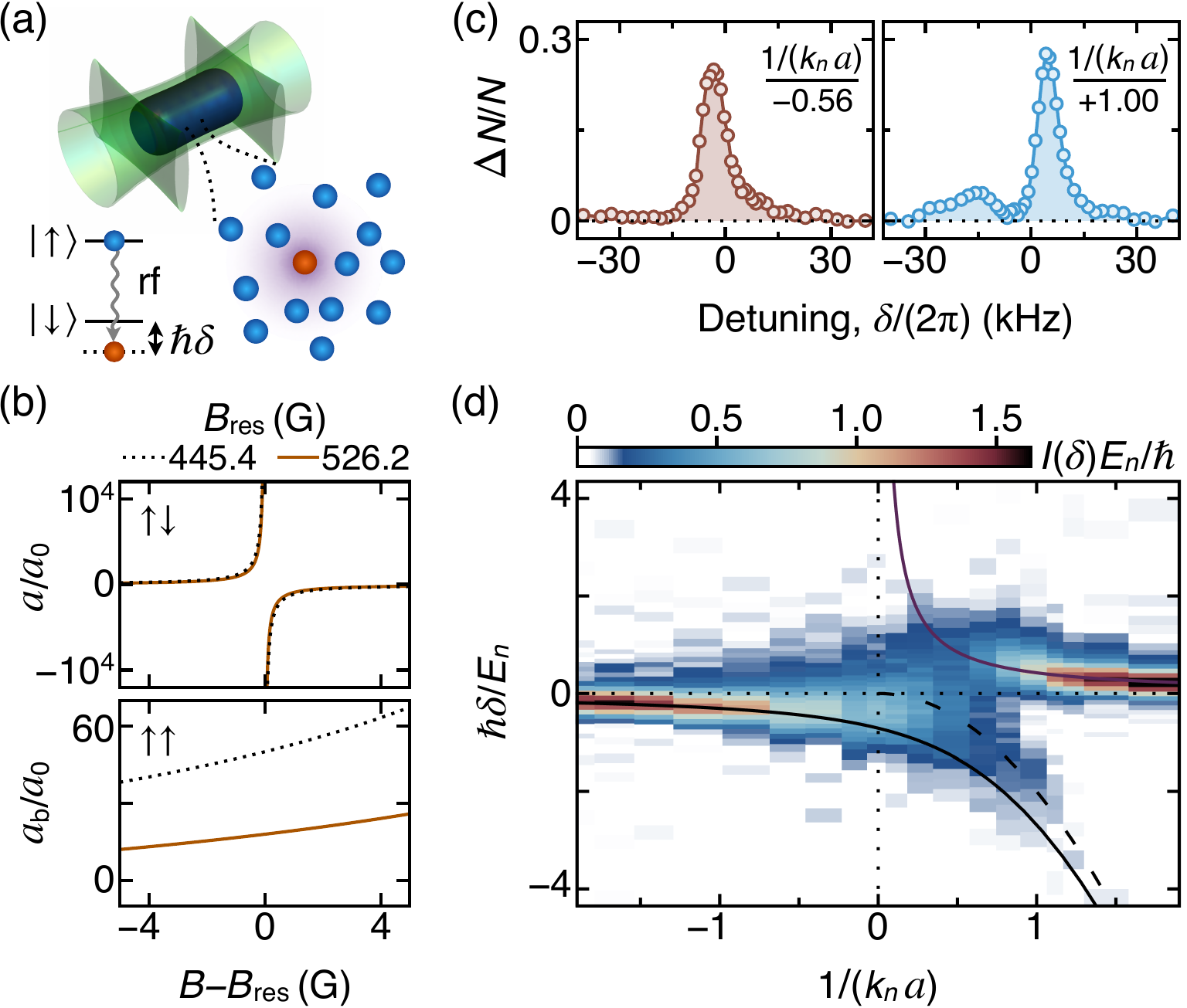}}
\caption{
Bose polarons in a box.
(a)~Sketch of the box trap and the polaron: an impurity (red, $\ket{\downarrow}$) interacts with a many-body bath (blue, $\ket{\uparrow}$) of homogeneous density $n$.
Starting with a spin-polarized BEC, we use an rf pulse (detuned by frequency $\delta/(2\pi)$ from the bare transition) to create a small population of the impurity state.
(b)~Impurity-bath interactions, characterized by the interstate scattering length~$a$, are tuned using one of two broad Feshbach resonances (top) that feature different values of the intrabath  
 scattering length $\ab$ (bottom); $a_0$ is the Bohr radius.
\mbox{(c,d)}~
Overview of the impurity spectrum near $\Bres= 526.2$\,G for $n\approx12\,\upmu$m$^{-3}$, corresponding to momentum scale $k_n\approx9\,\upmu$m$^{-1}$ and energy scale $E_n/\hbar \approx 2\pi \times 10$\,kHz.
(c)~Characteristic injection spectra (measured via fractional atom loss $\Delta N/N$, see text): for $a=-3900a_0$ (left) a single spectral feature is observed, while for $a=2200a_0$ (right) 
the spectrum is bimodal.
(d)~Injection spectra $I(\delta)$ across the Feshbach resonance, with all quantities expressed in dimensionless form. The solid lines show the single-phonon ansatz predictions~\cite{Rath:2013,Li:2014} for the energy of the attractive (black) and repulsive (purple) polaron. The dashed line shows the bare Feshbach dimer energy $\Ed$.
}
\label{fig1}
\vspace{-2em}
\end{figure}

\begin{figure*}[t!]
\centerline{\includegraphics[width=1.0\textwidth]{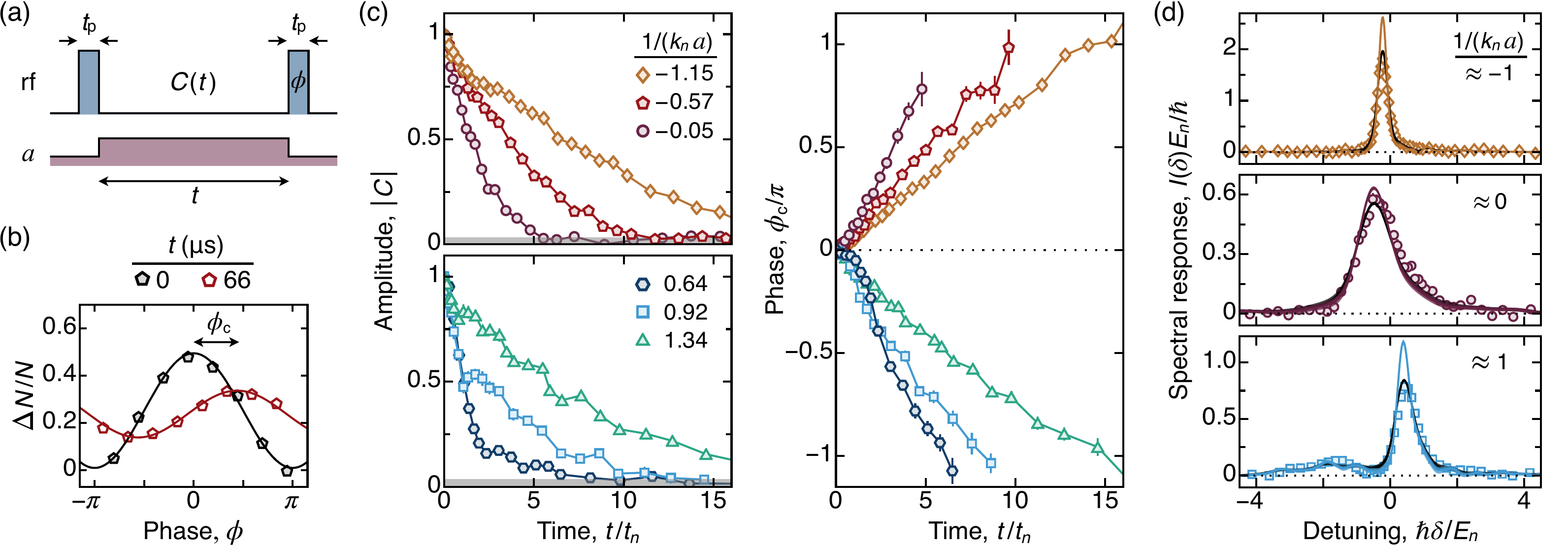}}
\caption{
Interferometric measurement of impurity dynamics.
(a)
We use two short rf pulses of length $t_{\rm p}$ separated by a variable time $t$ to access the coherence function $C(t) =|C(t)| \exp[i\phi_{\rm c} (t)]$.
We perform the pulses at weak interactions and use $B$-field quenches to set $a$ in between. 
(b) Typical traces of $\Delta N/N$ versus the phase of the second pulse, $\phi$, at $1/(k_n a)=-0.57$; $C(t)$ is obtained from the contrast and phase shifts using sinusoidal fits (solid lines).
(c) Evolution of $|C(t)|$ (left) and $\phi_{\rm c}(t)$ (right)
for varying $a$ at fixed $n=12(1)\upmu $m$^{-3}$ ($t_n\approx 16\,\upmu$s). 
The error bars reflect fitting errors and our noise floor for $|C|$ is $\approx3\%$ (shaded band).
(d)
Comparison of the interferometric and spectroscopic measurements, for three different interaction parameters. We show the Fourier transforms of $C(t)$ (colored lines), the injection spectra (symbols), and the former numerically broadened to mimic the finite-$\trf$ broadening of the latter (black lines). The line thickness and error bars reflect measurement errors.
}
\label{fig2}
\end{figure*}

Bose polarons in the strongly interacting regime have so far been investigated using harmonically trapped BECs~\cite{Hu:2016, Jorgensen:2016, Yan:2020, Skou:2021, Skou:2022, Morgen:2025}, with the measurements averaging their properties over an inhomogeneous bath density. Here we realize the textbook scenario of impurities injected into a quasi-uniform BEC, prepared in an optical box trap~\cite{Navon:2021}. 
We use different hyperfine states of $^{39}$K as the impurities and bath atoms, employ two broad Feshbach resonances to vary both $a$ and $\ab$, and also vary the homogeneous bath density $n$.
For a broad range of parameters, we observe consistent injection spectra and real-time dynamics, and find that they are universally set by the dimensionless interaction parameter $1/(k_n a)$ and the energy $E_n=\hbar^2 k_n^2/(4m_r)$, where \mbox{$k_n=(6\pi^2 n)^{1/3}$}, $m_r=m/2$, and $m$ is the atom mass. For $1/(k_n a)<0$, the impurity spectrum features a single branch; for weak interactions this corresponds to a well-defined attractive polaron, but near the resonance the spectra broaden dramatically, suggesting a breakdown of the quasiparticle picture.
Continuing across the resonance, up to $1/(k_n a)\approx0.5$ we still observe a single broad spectral feature, while for $1/(k_n a)\gtrsim0.5$ we resolve two separate spectral branches: the attractive branch that now peaks close to $\Ed$, but has a width $\propto E_n$ (revealing its many-body character), and a repulsive polaron branch.

\section{Experimental system}
Our experiments start with a quasi-pure weakly interacting $^{39}$K BEC, confined in an optical box trap~\cite{Gaunt:2013,Eigen:2016,Navon:2021} and spin-polarized in a hyperfine state denoted $\ket{\uparrow}$.
A small population of impurities in an adjacent hyperfine state $\ket{\downarrow}$ can be created using an rf pulse [see Fig.~\ref{fig1}(a)].
We either perform injection spectroscopy, measuring the fraction of $\ket{\downarrow}$ atoms after a pulse of duration $\trf$ and varying frequency $\omega/(2 \pi)$, or interferometry based on two short rf pulses separated by an evolution time $t$~\cite{Vale:2021}. We work with two combinations of $\ket{\uparrow}$ and $\ket{\downarrow}$ states, near two different broad Feshbach resonances [see Fig.~\ref{fig1}(b)]; in the low-field $|F,m_F\rangle$ basis, we use either $\ket{\uparrow}=|1,-1\rangle$ and $\ket{\downarrow}=|1,0\rangle$ with $\Bres=526.16(3)\,$G, or 
$\ket{\uparrow}=|1,0\rangle$ and $\ket{\downarrow}=|1,1\rangle$ with $\Bres=445.42(3)\,$G~\cite{Etrych:2023,bres}. The two bath states have $\ab$ values that differ by a factor of $\approx 3$ near $\Bres$,
and we vary $n$ in the range $(3-22)\,\upmu$m$^{-3}$ [corresponding to $E_n/\hbar = 2\pi \times (4-15)$\,kHz and characteristic time $t_n=\hbar/E_n=(10-40)\,\upmu$s], such that the intrabath interaction parameter $k_n \ab$ varies in the range $(5-28)\times 10^{-3}$. We calibrate $n$ by measuring the impurity mean-field energy at low $|a|$~(see Appendices~A-D for details). 

For large $|a|$, the recombination loss of particles (in both states) is fast compared to our measurement time and makes a direct measurement of the $\ket{\downarrow}$ population challenging. Instead, at the end of each experimental sequence we quench $B \rightarrow \Bres$ and wait for all impurities to be lost, so their concentration is faithfully reflected in the fractional loss of the total atom population, $\Delta N/N$ (see Appendix B).

\vspace{-1em}
\section{Injection spectroscopy}
In order for the measured injection spectrum, $I(\omega) \propto \Delta N(\omega)/N$, to reflect the energy spectrum of the impurity-bath system, we use long pulses $t_{\rm rf} \geq 200\,\upmu$s$\,\gg t_n$ and limit the injection fractions to $\lesssim10\%$, to minimize Fourier broadening and stay in the linear response regime.

Figure~\ref{fig1}(c) shows characteristic injection spectra in the strongly interacting regime, plotted as a function of the detuning $\delta=\omega_0-\omega$, where $\omega_0/(2\pi)$ (of about $0.1$\,GHz) is the bare transition frequency. For $1/(k_n a)=-0.56$ (left), we observe a single asymmetric spectral feature peaked at negative~$\delta$, corresponding to the attractive polaron. For $1/(k_n a)=1.00$ (right), we resolve two spectral features, corresponding to the attractive branch at $\delta<0$ and the repulsive polaron at $\delta>0$.

In Fig.~\ref{fig1}(d) we show an overview of $I(\delta)$ for a broad range of $1/(k_n a)$, with $n\approx12\,\upmu$m$^{-3}$, across the Feshbach resonance at $\Bres=526.2$\,G; here $t_{\rm rf}=200\,\upmu$s and the rf Rabi frequency is $\Omega/(2\pi)\approx 0.6$\,kHz. The black solid line shows the energy of the attractive polaron calculated using a minimal variational model~\cite{Chevy:2006b}, which includes single-phonon excitations of the BEC~\cite{Rath:2013,Li:2014}.
Within this theory, the attractive polaron is the ground state, and its energy approaches $\Ed$ (dashed line) at $1/(k_n a)\gg 1$, while the repulsive polaron is metastable and vanishes near the resonance; in the region where it is well-defined, the repulsive polaron has approximately the mean-field energy $2\pi \hbar^2 n a/m_r$ (purple line).

\vspace{-1em}
\section{Real-time dynamics}

To probe the real-time impurity dynamics, we use Ramsey-type interferometry~\cite{Cetina:2016,Fletcher:2017,Skou:2021} outlined in Fig.~\ref{fig2}(a), which gives the complex coherence function $C(t)=|C(t)|\exp[i\phi_{\rm c}(t)]$, formally related to $I(\delta)$ by a Fourier transform~\cite{Shashi:2014}. 
Here the first rf pulse (of duration $t_{\rm p} \approx 15\,\upmu$s) creates a small coherent admixture of impurities, and the second one, with a variable phase $\phi$, probes their evolution; the contrast and phase of the periodic variation of $\Delta N/N$ with $\phi$ [see Fig.~\ref{fig2}(b)] give $|C(t)|$ and $\phi_{\rm c}(t)$, defined so that $C(0)=1$. 
To resolve ultrafast dynamics (on $t_n$ timescale) and avoid effects of finite $t_{\rm p}$, we perform the two rf pulses at weak interactions ($a = \pm 440a_0$) and in-between set $a$ (in~$\sim2\,\upmu$s$\,\ll t_n$) for the evolution time~$t$ using magnetic-field quenches~\cite{Eigen:2017,windingfn}, without crossing $\Bres$.

In Fig.~\ref{fig2}(c) we show $C(t)$ for various $1/(k_n a)$ at fixed $n=12(1)\upmu$m$^{-3}$, for $\Bres=526.2\,$G. Both the phase winding and the contrast decay are generally faster for stronger interactions. 
However: (i) for $1/(k_n a)<0$, the dynamics are essentially independent of $a$ for $t/t_n \ll 1$, as previously studied in detail in Refs.~\cite{Skou:2021,Skou:2022} (see also Appendix~E), and (ii)~for $1/(k_n a)>0.5$ the initial dynamics are complicated because of the beating between the two branches~\cite{Cetina:2016,Morgen:2025}, but the negative-energy (attractive) component decoheres faster and the late-time dynamics are dominated by the repulsive polaron; note that in our definition of $C(t)$, positive energy corresponds to a negative slope of $\phi_{\rm c}(t)$.

In Fig.~\ref{fig2}(d) we show, for three characteristic $1/(k_n a)$, that we obtain essentially the same results from interferometry and spectroscopy, except that interferometry is intrinsically free of Fourier broadening. Here one also sees that for $1/(k_n a) \approx1$ the negative-energy branch is broader than the positive-energy one, contradicting the often-made assumption that the latter is the broader of the two.

\section{
Universal dynamics at unitarity}

\begin{figure}[t!]
\centerline{\includegraphics[width=1\columnwidth]{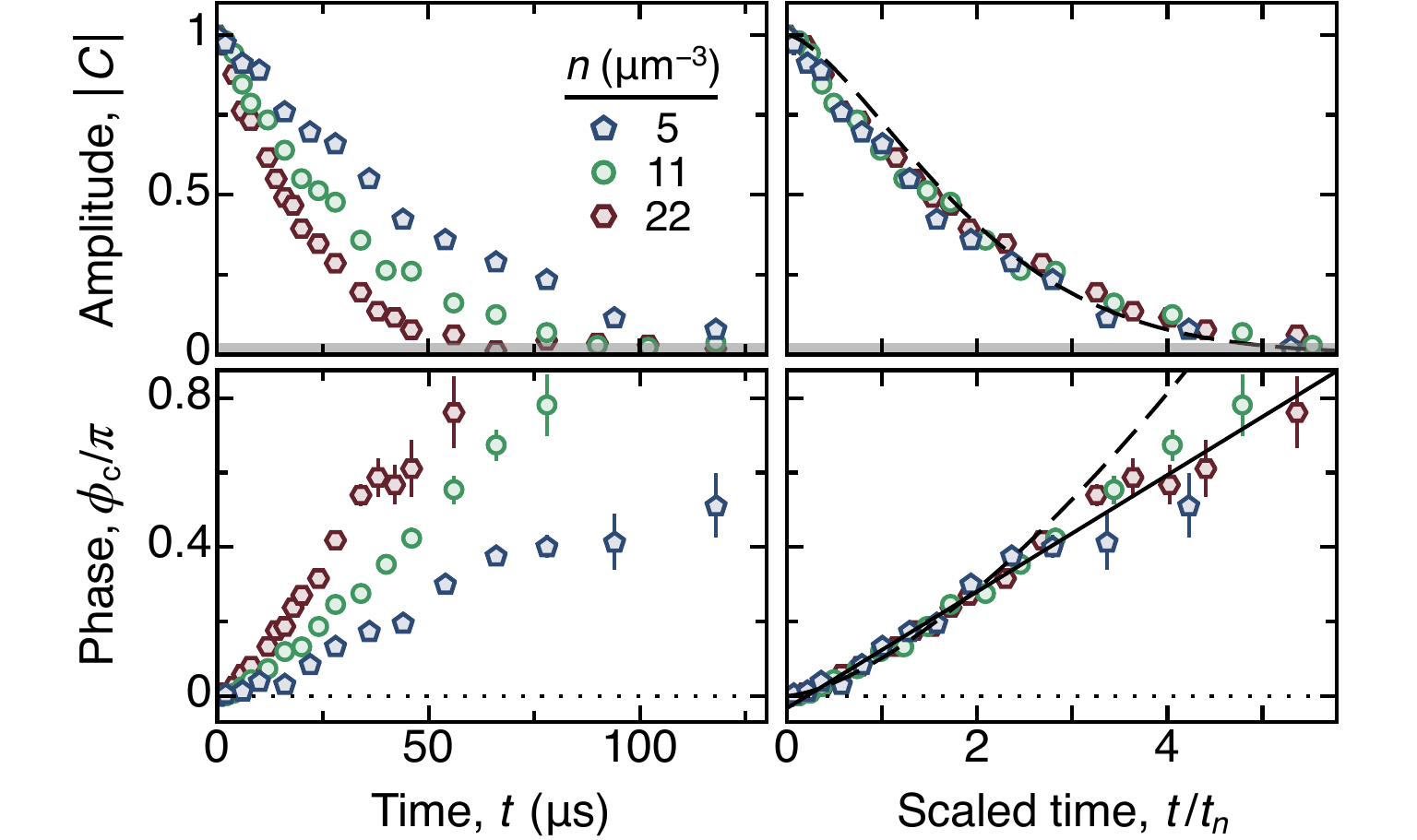}}
\caption{
Universal dynamics at unitarity.
Evolution of the coherence-function amplitude $|C|$ (top) and phase $\phi_{\rm c}$ (bottom) for three different bath densities $n$.
The left panels show the evolution with respect to the real time $t$; for larger $n$ the dynamics are faster.
In the right panels, the same data collapse onto universal curves when plotted versus the scaled time $t/t_n$.
The dashed lines show the predictions of~\cite{Drescher:2021} [Eq.~(\ref{eq:ideal}), see text for details], where the polaron is not well-defined at unitarity.
The solid line shows a linear fit of $\phi_{\rm c}$ for $t/t_n>1$. The error bars show fitting errors.
}
\label{fig3}
\end{figure}

We now turn to a detailed study of the impurity dynamics at unitarity, where $a$ diverges and drops out of the problem. In Fig.~\ref{fig3}, we show $C(t)$ for $\Bres=526.2\,$G and three different $n$. The dynamics are naturally faster for larger $n$ (left), but when plotting versus $t/t_n$, both $|C|$ and $\phi_{\rm c}$ data collapse onto universal curves (right). This `Fermi-like' scaling with $t_n\propto n^{-2/3}$ means that the physics is universal and scale-invariant (solely set by $n$, which defines the absolute time and energy scales); see also Fig.~\ref{fig12} in Appendix~G. Such scaling is a hallmark result for unitary single-component Bose gases~\cite{Ho:2004a,Klauss:2017,Eigen:2017,Eigen:2018,bosefermiloss}, and here we observe it for the first time for an impurity interacting with a BEC.

Without resorting to any model-dependent theory, we first note that $|C|$ decays to $1/e$ within about $2t_n$, and during this time $\phi_{\rm c}$ winds only by about $1~{\rm rad}$.
The fact that the decoherence and phase-winding rate are essentially the same ($\approx 0.5/t_n$) suggests that at unitarity the polaron quasiparticle is at best marginally defined; note that the decay of $|C|$ also has a contribution from particle loss, but this is significantly smaller ($\approx 0.1/t_n$) than the lossless decoherence rate (see Appendix~F).

More quantitatively, we find (surprisingly) good agreement with the zero-temperature theory of Ref.~\cite{Drescher:2021} (see also~\cite{Ardila:2021}) for an `ideal Bose polaron' at unitarity, where $C(t)$ has the analytical form:
\be
\begin{split}
  |C(t)|&=\exp\left(-\frac{16}{9\pi^{3/2}}\left(\frac{t}{t_n}\right)^{3/2}\right),\\
  \phi_{\rm c}(t)&= \frac{16}{9\pi^{3/2}}\left(\frac{t}{t_n}\right)^{3/2} \, .
\end{split}
\label{eq:ideal}
\ee
In this idealized theory the impurity mass is infinite (so $m_r=m$) and $\ab=0$, which leads to the orthogonality catastrophe (no overlap of the ground-state wave functions with and without an impurity), the polaron quasiparticle is not well-defined (see also \cite{Yoshida:2018,Massignan:2021,Guenther:2021}), and the dynamics are instead dominated by excited states of the system; to compare to this theory we just set $m_r$ to our value $m/2$.
The theory (dashed lines in Fig.~\ref{fig3}) captures $|C|$ reasonably well over our full time range, and it also captures $\phi_{\rm c}$ up to about $2t_n$, while at longer times the data is captured better by a linear fit (solid line) with a slope of $0.49(4)/t_n$. This phase winding is significantly slower than that predicted for the ground-state polaron, which is typically faster than $1/t_n$~(see e.g.\cite{Ardila:2015,Yoshida:2018}).
While these ground-state predictions could be wrong, a more likely explanation is that the dynamics are dominated by the excited states.

A similarly fast decay of $|C|$ on the timescale of order $t_n$, or equivalently spectral broadening on the scale of $E_n/\hbar$, was previously observed in harmonic-trap experiments~\cite{Hu:2016,Jorgensen:2016,Skou:2021,Skou:2022}, but it could largely be explained by inhomogeneous broadening, even if one assumed that locally (at a fixed density) the spectrum was sharp~\cite{Ardila:2019,Skou:2021,Skou:2022}. Our measurements show that the fast decay of $|C|$ at unitarity is in fact an intrinsic homogeneous effect.

\begin{figure*}[t!]
\centerline{\includegraphics[width=1\textwidth]{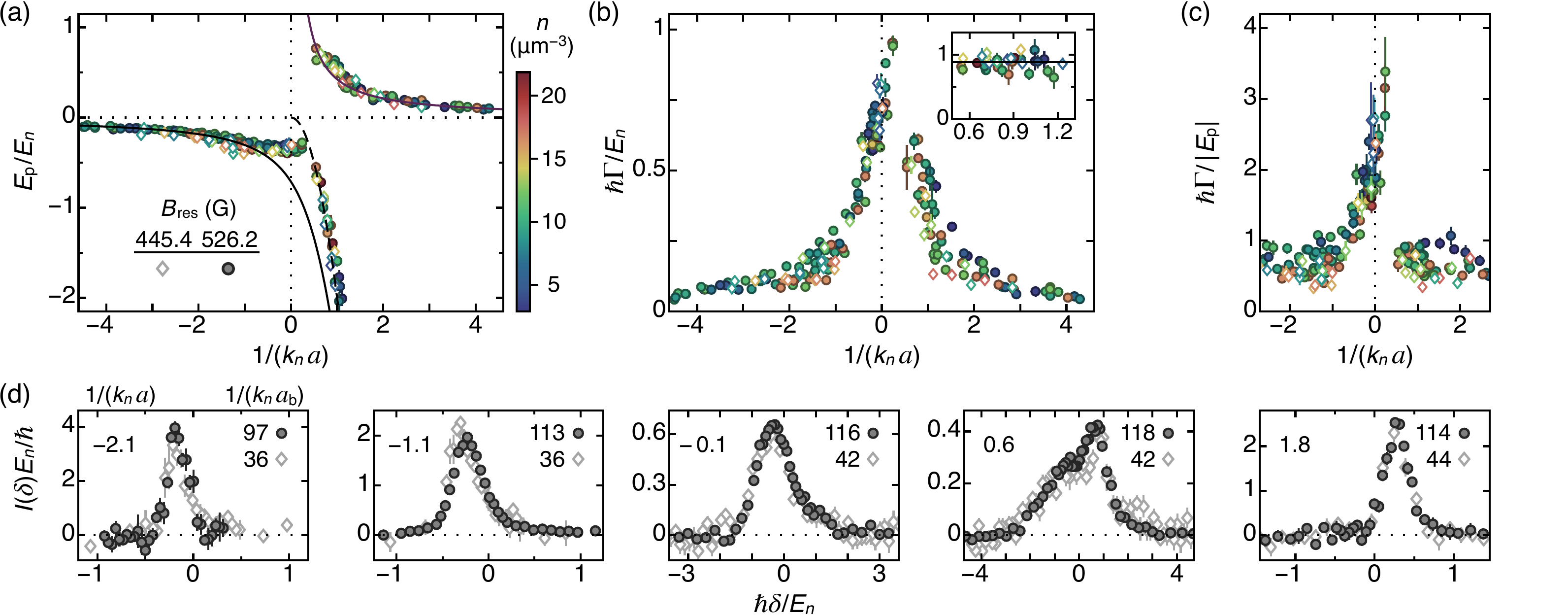}}
\caption{
Universal features of the Bose polaron spectrum.
(a) Spectrum mode $\Ep$, and (b) half-width at half-maximum $\hbar\Gamma$~\cite{residualtrf},
for various densities $n$ (color bar) and two different Feshbach resonances (legend). When normalized by $E_n$ and plotted versus the dimensionless $1/(k_n a)$, all data essentially collapse onto universal curves.
In (a), the solid and dashed lines are the same as in Fig.~\ref{fig1}(d).
In (b), for $1/(k_n a)\gtrsim 0.5$, where we resolve two branches, the main panel and inset show, respectively, data for the repulsive polaron and the attractive branch.
The solid line in the inset shows the average value, $\hbar \Gamma/E_n =0.88$. 
(c) Ratio of $\hbar\Gamma$ and $|\Ep|$ for the repulsive branch and the attractive one for $1/(k_n a)\lesssim0.3$.
Near resonance, $\hbar\Gamma$ exceeds $|\Ep|$ for the attractive polaron, signaling a breakdown of the quasiparticle picture.
(d)~Comparison of the full spectra for different $1/(k_n \ab)$. Each panel shows spectra at a fixed $1/(k_n a)$ and two values of $1/(k_n \ab)$.
The error bars reflect fitting errors (a-c) and measurement errors (d).}
\vspace{0.5em}
\label{fig4}
\end{figure*}

\vspace{-1em}
\section{
Universality of the Bose polaron spectrum}
\vspace{-0.5em}
We now extend our study to a broad range of 
$n$ and $a$, and to the Feshbach resonance at $\Bres=445.4\,$G, to also vary~$\ab$.
In Fig{s.~\ref{fig4}(a-c), we characterize the impurity spectra using their mode, $\Ep/\hbar$, and half width at half maximum, $\Gamma$~\cite{residualtrf}; here we employ the more economical spectroscopic measurements (adjusting $\trf$ and $\Omega$ to minimize the Fourier broadening), and for $1/(k_n a)>0.5$ we extract $\Ep$ and $\Gamma$ for the two resolved branches separately~(see Appendix~B).

As shown in Fig.~\ref{fig4}(a), plotting $\Ep/E_n$ versus $1/(k_n a)$ collapses almost all our data, and only for $1/(k_n a)\approx -1.5$ we see hints of a small difference between the two resonances. 
This implies that the physics depends on $n$ and $a$ primarily through the dimensionless $k_n a$, and moreover that any additional scales (including $\ab$) enter only weakly~\cite{extralengthscales}.

For the attractive branch, $\Ep/E_n$ is consistent with the single-phonon ansatz (black line) for $1/(k_n a)\lesssim-2$, but for stronger attractive interactions it deviates and tends to a constant near unitarity, while for $1/(k_n a)\gtrsim 0.5$, $\Ep$ is close to $\Ed$ (dashed line).
For the repulsive polaron branch, $\Ep$ is close to the mean-field prediction (purple line).

Figure~\ref{fig4}(b) shows the dimensionless $\hbar\Gamma/E_n$, which is also predominantly set by $1/(k_n a)$, albeit with a larger data scatter than for $\Ep/E_n$ in Fig.~\ref{fig4}(a).
Surprisingly, for the attractive branch at $1/(k_n a)\gtrsim 0.5$ (see inset), $\hbar \Gamma / E_n$ is essentially constant, meaning that $\Gamma$ is independent of $a$.

In Fig.~\ref{fig4}(c), we plot the ratio of $\hbar\Gamma$ and $\Ep$ for the repulsive branch and the attractive one for $1/(k_n a)\lesssim0.3$.
For the attractive polaron, $\hbar\Gamma/|\Ep|$ rises above unity near resonance, again suggesting a breakdown of the quasiparticle picture. 
The repulsive polaron vanishes for $1/(k_n a)\lesssim 0.5$, but at all $1/(k_n a)$ where we can resolve it, $\hbar\Gamma/|\Ep|$ is below unity. For both polaron branches, away from resonance $\hbar \Gamma/\Ep$ is approximately constant, which can be partially explained by finite-size effects seen in mean-field simulations~(see Appendix~D)~\cite{Epcomment}. 

Going beyond the characterization of the spectra by $\Ep$ and $\Gamma$, in Fig.~\ref{fig4}(d) we show for select $k_n a$ values that the full spectral shapes are the same for fixed $k_n a$ and different $\ab$.

\begin{figure}[t!]
\centerline{\includegraphics[width=1.\columnwidth]{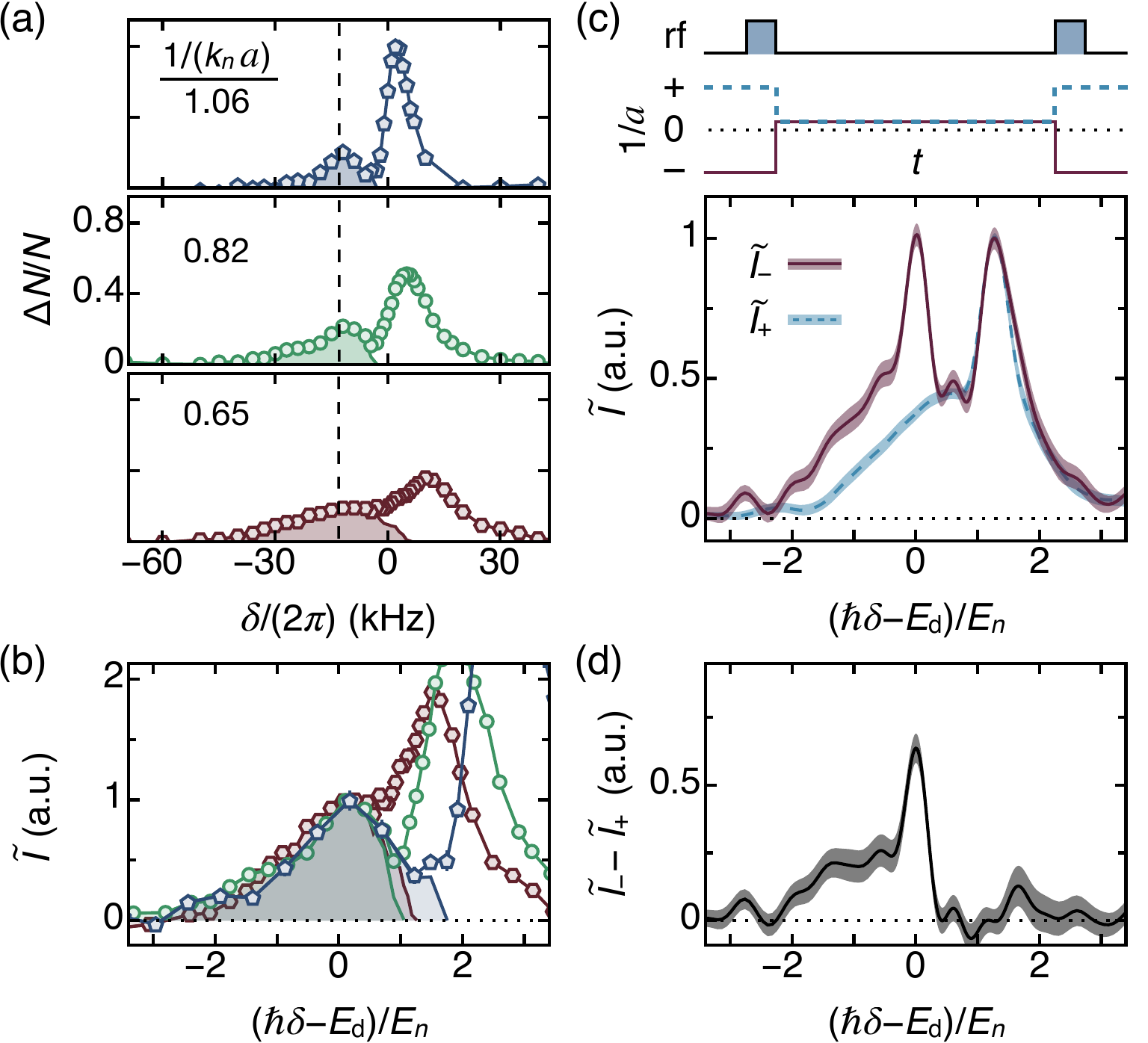}}
\caption{
The attractive branch at $1/(k_na)\gtrsim 0.5$.
(a) Injection spectra for varying $n$ at $a=2700a_0$, $\Ed/(2\pi \hbar)=-13$\,kHz (dashed line). 
The shaded attractive-branch contribution is obtained by subtracting a fit to the repulsive polaron~(see Appendix~E).
(b) When normalized by its peak and plotted versus $(\hbar\delta-\Ed)/E_n$, the attractive part of $I(\delta)$ collapses onto a universal asymmetric curve.
(c) Differential interferometry. (Top) We perform interaction quenches starting from either weak repulsive or weak attractive interactions to the same $1/(k_n a)$, to obtain spectra with different relative weights of the two branches.
(Bottom) The resulting spectra $\tilde{I}_-$ and $\tilde{I}_+$, normalized by the height of the repulsive polaron peak; here $1/(k_n a)\approx0.6$ and $n\approx13\,\upmu$m$^{-3}$.
(d)~The difference $\tilde{I}_- - \tilde{I}_+$ reveals
the attractive part of the spectrum, with the same peak and asymmetry as in (b).
The error bars and line thicknesses reflect measurement uncertainties.
}
\label{fig5}
\vspace{-1em}
\end{figure}

In Appendix~G we show three further figures complementing Fig.~\ref{fig4}.
In Fig.~\ref{fig13} we show a comparison of the real-time dynamics at different $n$ and the same $\ab$.
In Fig.~\ref{fig14} we show the generally good agreement between our $\Ep$ and $\Gamma$ values obtained from spectroscopy and interferometry. Finally, in Fig.~\ref{fig15} we compare our data with the injection measurements performed in harmonic traps~\cite{Hu:2016,Jorgensen:2016,Morgen:2025}, which all used broad Feshbach resonances but
explored different gas densities, $\ab$ values, and ratios of the impurity and bath-atom masses. While spectral widths are not directly comparable due to inhomogeneous broadening and other technical differences, we find that $\Ep/E_n$ is consistent across experiments when $E_n$ is defined using the mean probed density and the appropriate $m_r$. This corroborates the universality seen in Fig.~\ref{fig4} over a much larger parameter range, spanning $E_n/\hbar=2\pi \times (4-70)$\,kHz and $\ab=(9-100)a_0$, but note that in all cases $k_n \ab= (5-110) \times 10^{-3}$ is relatively small.

\vspace{-1em}

\section{The attractive branch at $1/(k_na)\gtrsim0.5$}
\vspace{-0.5em}
Finally, we further investigate the attractive branch for $1/(k_n a)\gtrsim0.5$.
In Fig.~\ref{fig5}(a), we show spectra for increasing $n$ at fixed \mbox{$\Ed/(2\pi \hbar)=-13$\,kHz} ($a=2700a_0$), for $\Bres=526.2\,$G~\cite{strongrf}.
This explicitly shows how the attractive part of the spectrum broadens with $n$, while its peak stays roughly at $\Ed/\hbar$ [see also Figs.~\ref{fig4}(a,b)].
As shown in Fig.~\ref{fig5}(b),
when normalized by their height and plotted versus $(\hbar\delta-\Ed)/E_n$, the attractive parts of the spectra collapse onto a universal curve; this is consistent with the $E_n$-set width shown in the inset of Fig.~\ref{fig4}(b).
This universal spectrum is dominated by states near $\Ed$; such excited states are predicted in variational theories that include at least two phonons~\cite{Jorgensen:2016}.
On the other hand, the asymmetric lineshape (shaded area), with a significant tail towards $\hbar\delta<\Ed$, suggests the existence of many-body state(s) with attractive-polaron character (with energy approximately $E_n$ below $\Ed$). This notion is also supported by the fact that in Fig.~\ref{fig1}(d) the lower edge of the spectrum appears to continuously connect across the resonance to the attractive polaron. However, understanding the full nature of this branch remains a challenge.

To corroborate these observations, we introduce `differential interferometry' outlined in Fig.~\ref{fig5}(c),
which allows us to isolate the attractive part of the spectrum in the regime where it overlaps with the repulsive-polaron one.
We perform two interferometry measurements at the same $1/(k_n a)\approx0.6$, but with the rf pulses performed at either weak attractive or weak repulsive interactions 
($a=\pm440a_0$). 
The two protocols result in different relative weights of the two branches: when quenching from $a<0$, across $\Bres$, the attractive-branch weight is significantly enhanced compared to a quench from $a>0$~\cite{iniaeffect}.
We normalize the two spectra so that they have the same height of the repulsive-polaron peak, and their difference [see Fig.~\ref{fig5}(d)] reveals the (unnormalized) attractive-branch contribution, which shows the same peak and asymmetric shape as in Fig.~\ref{fig5}(b).

\vspace{-1.5em}
\section{Conclusion and outlook}
\vspace{-0.5em}
We performed a comprehensive study of the dynamics and spectral features of impurities strongly interacting with a homogeneous BEC.
For our case of broad Feshbach resonances, equal masses of impurities and bath atoms, and weak intra-bath interactions, we reveal remarkably universal behavior set by the energy $E_n$, defined by the bath density, and the single dimensionless interaction parameter $1/(k_n a)$. 
Our measurements near unitarity indicate a breakdown of the quasiparticle picture, qualitatively consistent with the bosonic orthogonality catastrophe (OC)~\cite{Guenther:2021,Yoshida:2018}.
However, the key prediction of the OC theories, namely the dependence of the quasiparticle residue on the system parameters, cannot be quantitatively tested in the current experiments because, unlike in Fermi systems~\cite{Kohstall:2012,Scazza:2017,Darkwah:2019,Adlong:2020}, the timescales for quasiparticle formation and decoherence are not clearly separated; devising methods to determine the residue thus remains an important challenge.
It would also be interesting to extend our study to narrow Feshbach resonances and to different mass ratios, where additional lengthscales could become relevant and break the observed universality.
Our system and the ability to resolve the two branches on the repulsive side of the resonance are also promising for future studies of mediated interactions between polarons~\cite{Camacho-Guardian:2018a,Camacho-Guardian:2018b} and the effects of nonzero temperature~\cite{Guenther:2018,Field:2020,Pascual:2021,Drescher:2024}. 

\vspace{-1em}
\section*{Acknowledgements}

We thank Martin Gazo, Georg Bruun, Martin Zwierlein, Richard Schmidt, Arthur Christianen, Tilman Enss, Moritz Drescher, Meera Parish, Jesper Levinsen, Fabian Grusdt, and Eugene Demler for discussions, and Jan Arlt, Andreas Morgen, Eric Cornell, and Ming-Guang Hu for sharing their data.
This work was supported by EPSRC [Grant No.~EP/P009565/1], ERC [UniFlat], and STFC [Grants No.~ST/T006056/1 and No.~ST/Y004469/1]. Z.H. acknowledges support from the Royal Society Wolfson Fellowship. C.E. acknowledges support from Jesus College (Cambridge). 

\vspace{-1em}
\section*{Author contributions}
J.E., G.M., A.C., and C.E. conceptualized the experiments and collected the data. J.E. and C.E. analyzed the data. All authors contributed significantly to the experimental setup, the interpretation of the results, and the production of the manuscript. Z.H. and C.E. supervised the project.

\vspace{-1em}
\section*{Data availability}
The data that support the findings of this article are openly available~\cite{repository}.

\setcounter{section}{0} 

\vspace{-1em}
\setcounter{equation}{0} 
\renewcommand\theequation{A\arabic{equation}} 
\section*{Appendix~A:~Preparation and measurement}

\begin{figure}[b!]
\centerline{\includegraphics[width=1\columnwidth]{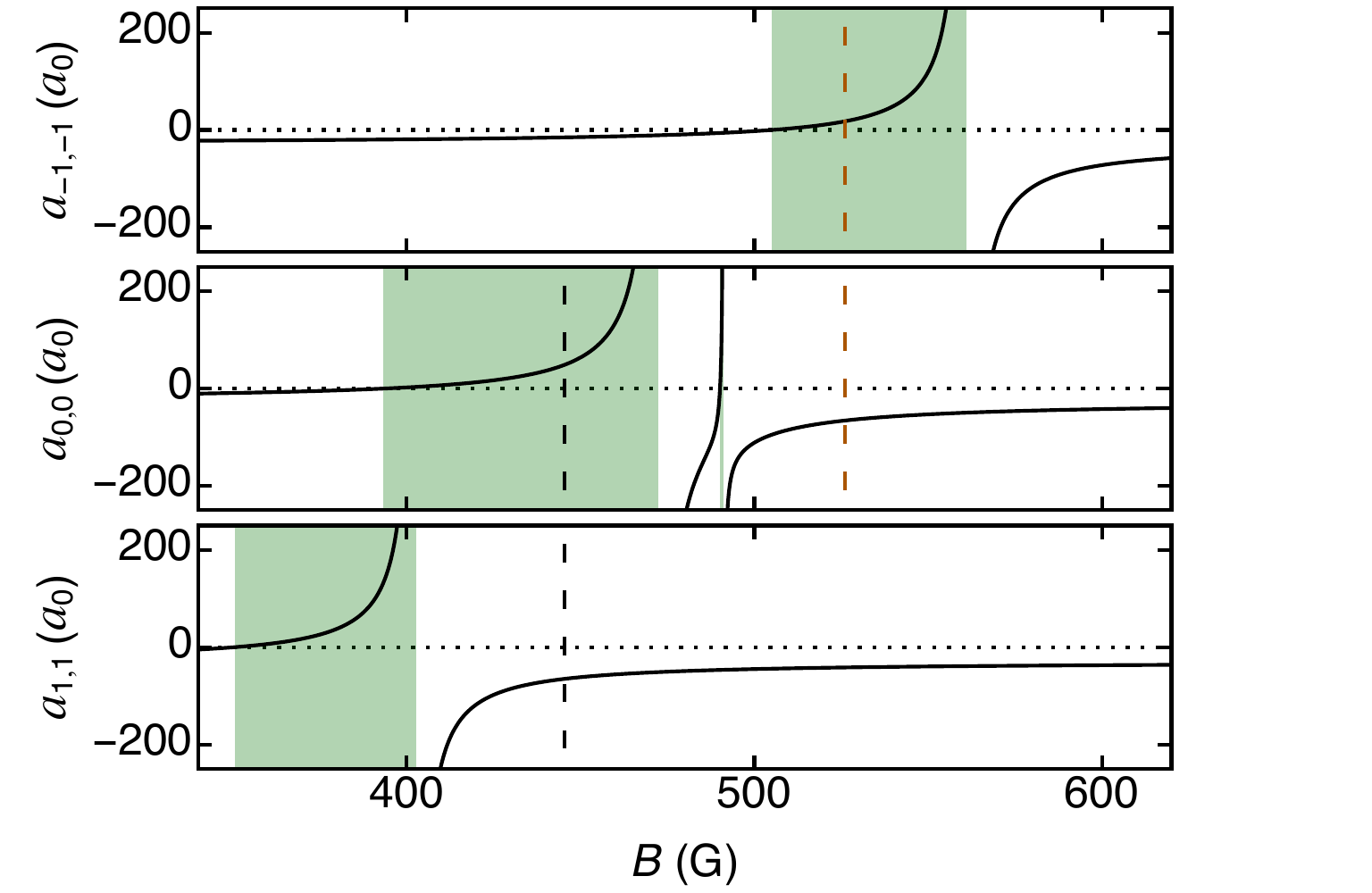}}
\caption{
Intrastate Feshbach resonance landscape for $^{39}$K in the \mbox{$F=1$} manifold\cite{Etrych:2023}; here we denote the different scattering lengths $a_{m_F,m_F}$. 
The dashed lines indicate $\Bres$ of the two interstate Feshbach resonances used to tune impurity-bath interactions.
The green shaded areas highlight regions where $a_{m_F,m_F}>0$. 
}
\label{fig6}
\end{figure}

\begin{figure*}[t!]
\centerline{\includegraphics[width=1\textwidth]{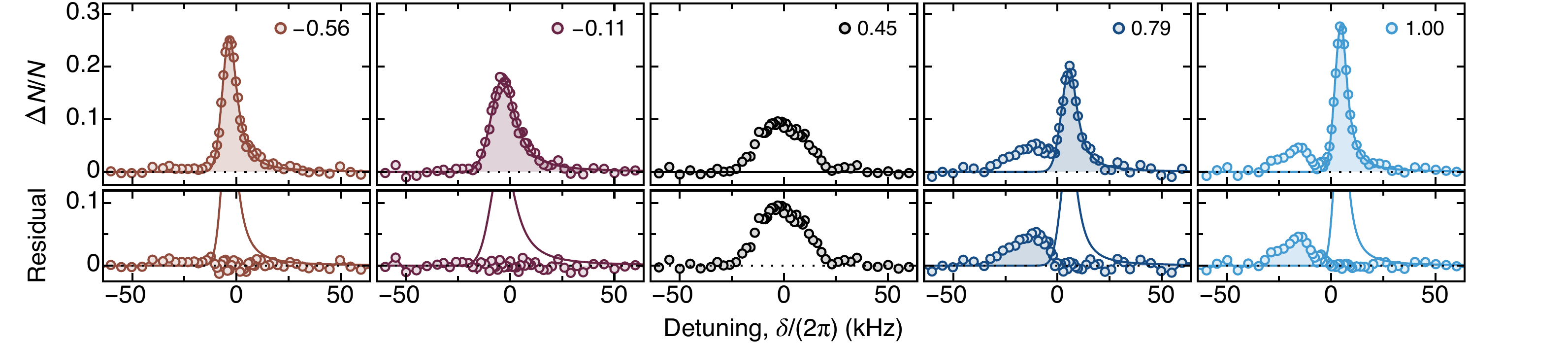}}
\caption{
Extraction of $\Ep$ and $\Gamma$.
(Top) Examples of $I(\delta)$ at different $1/(k_n a)$ (legends) and the polaron spectrum fits (see text).
(Bottom)~Subtraction of the polaron fit reveals the molecular spectrum for $1/(k_n a)\gtrsim0.5$. At $1/(k_n a)=0.45$, the two branches cannot be resolved.
}
\label{fig7}
\vspace{-1em}
\end{figure*}

Our optical box trap is cylindrical (radius $R$, length $L$), and to increase the range of densities we can explore, we use either $R\approx14\,\upmu$m and $L\approx45\,\upmu$m or $R\approx10\,\upmu$m and $L\approx30\,\upmu$m, with a trap depth $\UD\approx\kB \times 50$\,nK.
Our experiments always start with a quasi-pure spin-polarized BEC in $\ket{F,m_F}=\ket{1,1}$, and we then transfer the atoms to either $\ket{1,0}$ or $\ket{1,-1}$ (our two bath states) using Landau--Zener rf sweeps.

For preparing the $\ket{1,0}$ bath, we transfer the BEC from $\ket{1,1}$ to $\ket{1,0}$ at $B=396$\,G, where both intrastate scattering lengths,  $a_{1,1}$ and $a_{0,0}$, are positive (see Fig.~\ref{fig6}).
For preparing the $\ket{1,-1}$ bath, we cannot simply transfer the BEC from $\ket{1,0}$ to $\ket{1,-1}$, because there is no $B$ at which both 
$a_{0,0}$ and $a_{-1,-1}$ are positive, and attractive interactions would lead to BEC collapse~\cite{Donley:2001,Eigen:2016}.
To circumvent this, at $B=396\,$G we temporarily create a highly nonthermal state with destroyed coherence but low energy per particle, by shaking the cloud with a periodic force~\cite{Martirosyan:2024}. This allows us to transfer the atoms to $\ket{1,-1}$ without collapse, then ramp $B$ to $\approx550\,$G where $a_{-1,-1}>0$, and recondense the cloud, achieving a final quasi-pure BEC with about $80\%$ of the initial $\ket{1,1}$ condensate atom number.

We levitate the bath atoms against gravity using a magnetic field gradient. Owing to the small difference in magnetic moments between the bath and impurity states ($3\%$ for $\ket{1,0}$ and $\ket{1,1}$ at $445\,$G, and $0.3\%$ for $\ket{1,-1}$ and $\ket{1,0}$ at $526\,$G), the impurities are also essentially levitated.

The impurity-impurity interactions are attractive with \mbox{$\ai\approx -64 a_0$} in both cases, but are not relevant on our experimental timescales due to the small impurity concentration.

At the end of the experimental sequence, we measure the total atom number after time of flight expansion (typically for $60$~ms) using absorption imaging at low $B$.
Our systematic atom-number uncertainty is $\approx 10\%$.
\vspace{-1em}
\setcounter{equation}{0} 
\renewcommand\theequation{B\arabic{equation}} 
\section*{Appendix~B:~Measuring impurity spectra}
\vspace{-1em}
In the linear response regime, injection spectroscopy reveals the impurity spectral function $A(\omega)$~\cite{Vale:2021, Scazza:2022}. For a square rf pulse of duration $\trf$, the spectrum is approximately:
\be
I(\omega)=\frac{\trf}{2\pi}\int_{-\infty}^\infty A(\omega^\prime)\, \mathrm{sinc}\left[\frac{(\omega-\omega^\prime)\trf}{2}\right]^2 \,{\rm d} \omega^\prime,
\label{Iconvolve}
\ee
normalized such that $\int I(\omega) \,{\rm d}\omega=1$.

For a weak pulse, the fraction of atoms transferred from $\ket{\uparrow}$ to $\ket{\downarrow}$ is $(\Omega^2 \trf \pi/2) I(\omega)$, where $\Omega$ is the Rabi frequency. In the experiment, we instead measure the total loss fraction (following a quench to $\Bres$), which is related to the transfer fraction by a constant of proportionality $\alpha$, so that $\Delta N/N= (\alpha \Omega^2 \trf \pi/2) I(\omega)$.
We experimentally estimate $\alpha \approx 4$ based on loss measurements at unitarity.

In Fig.~\ref{fig1}(d), we normalize the spectra using $\trf=200\,\upmu$s, \mbox{$\alpha=4$}, and estimated $\Omega/(2\pi)= 0.6$\,kHz. In Fig.~\ref{fig2}(d), we normalize all spectra individually based on their numerically estimated area.

The spectral function can also be calculated from the coherence function $C(t)$:
\be
A(\omega)=\frac{1}{\pi}\,\mathrm{Re}\left[\int^\infty_0 C(t) \exp({-i\omega t}) \,{\rm d} t\right].
\label{AfromC}
\ee
To numerically compute the Fourier transform, we use piece-wise linear interpolations of $|C|$ and $\phi_{\rm c}$. We assess the uncertainty in $A(\omega)$ by repeating the procedure with points randomly sampled within their experimental errors.
For comparison with spectroscopic data [Fig.~\ref{fig2}(d)], we can also take into account Fourier broadening using Eq.~(\ref{Iconvolve}). 

\begin{figure}[b!]
\centerline{\includegraphics[width=1\columnwidth]{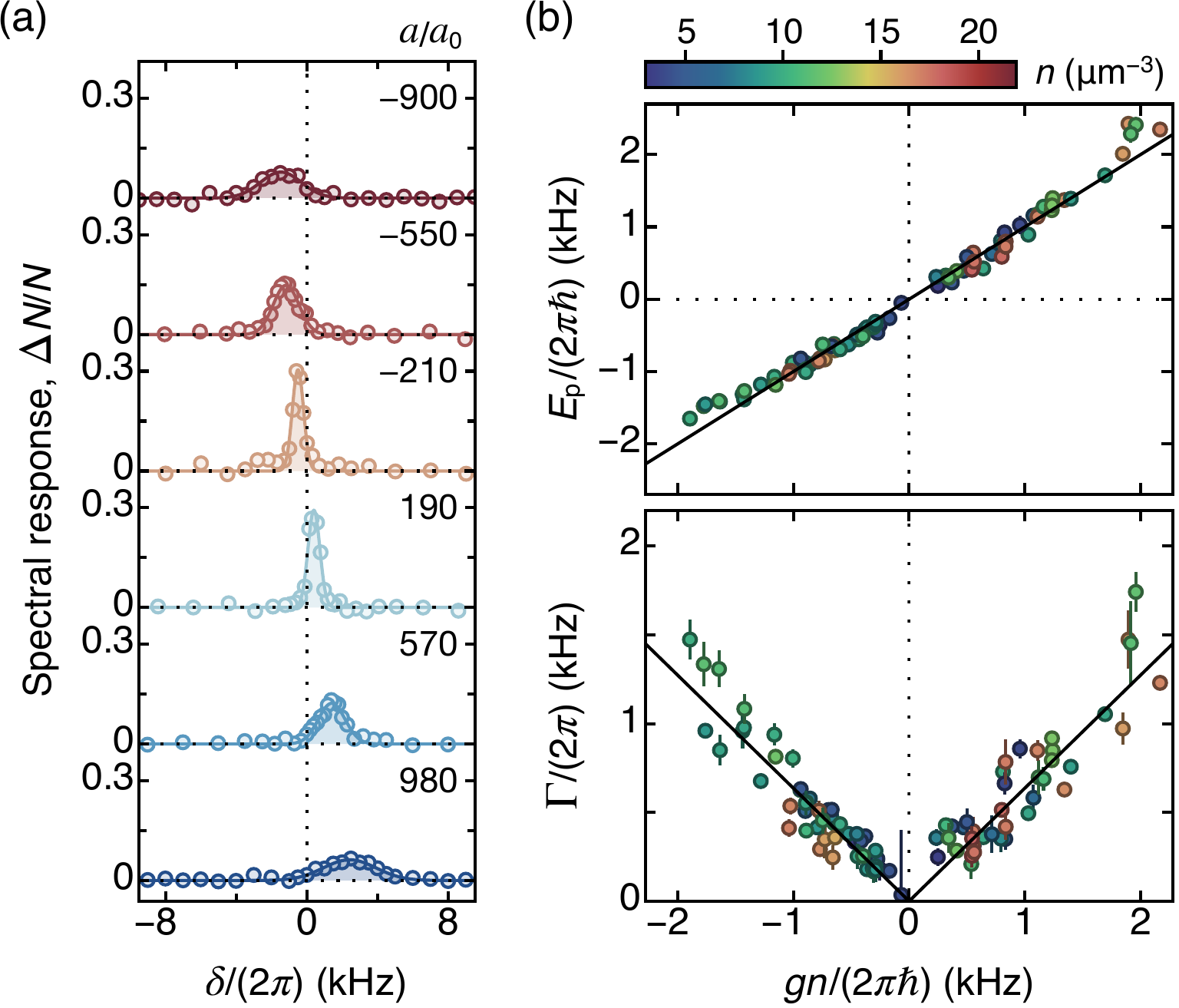}}
\caption{
Injection spectroscopy in the weakly interacting regime. (a) Injection spectra for $n=12(1)\,\upmu $m$^{-3}$ and varying $a$, for $\Bres=526.2\,$G; here $t_{\rm rf}=1600\,\upmu$s.
The solid lines show Gaussian fits used to extract the mode $E_{\rm p}$ and half-width-half-maximum $\Gamma$ \cite{residualtrf}.
(b)~Plot of $\Ep$ (top) and $\Gamma$ (bottom) versus $gn$ for a range of $n$ and $a$.
The solid lines show $\Ep = gn$ (top) and  $\Gamma=0.64|g|n/\hbar$ (bottom). The error bars reflect measurement errors (a) and fitting errors (b).
}
\label{fig8}
\end{figure}

Our experimental $B$-field stability is $\sim 10$~mG, corresponding to $\pm0.03$\,kHz and $\pm0.4$\,kHz uncertainty in $\omega_0/(2\pi)$ near $\Bres=526\,$G and $445$\,G, respectively. We also correct for the small shift in $\delta$ due to the mean-field energy of the bath atoms, $\delta_{\rm mf}= 4 \pi \hbar n\ab/m$, where $m$ is the atom mass.

In Fig.~\ref{fig7}, we exemplify how we extract the mode $\Ep$ and half-width $\Gamma$ of $I(\delta)$ for different characteristic $1/(k_n a)$ in the strongly interacting regime.
The polaron features are asymmetric, with a tail to larger $\delta$.
To extract $\Ep$ and $\Gamma$, we heuristically capture the spectra by fitting them using a combination of a Gaussian (for $\delta<\Ep/\hbar$) and a Lorentzian (for $\delta>\Ep/\hbar$).

In the cases where we resolve the repulsive polaron and the `molecular' branch, we first fit the repulsive polaron spectrum over a constrained range, and then subtract this fit to obtain the molecular contribution to the spectrum. 
We refrain from extracting $\Ep$ and $\Gamma$ between \mbox{$0.3\lesssim1/(k_n a)\lesssim0.5$} (see example at $1/(k_n a)=0.45$ in Fig.~\ref{fig7}).

We always correct $\Gamma$ extracted from injection spectroscopy for Fourier broadening using \mbox{$\Gamma=\left(\Gamma_{\rm e}^2-\Gamma_{\rm t}^2\right)^{1/2}$, where $\Gamma_{\rm e}$} is the raw extracted width and \mbox{$\Gamma_{\rm t}\approx2.78/\trf$} is the width of the response function in Eq.~(\ref{Iconvolve}).

\vspace{-1em}
\setcounter{equation}{0} 
\renewcommand\theequation{C\arabic{equation}} 
\section*{Appendix~C:~
Density calibration}
\vspace{-1em}
In Fig.~\ref{fig8} we show injection spectra for $1/(k_n |a|)>2$.
As shown in Fig.~\ref{fig8}(a), we observe essentially symmetric spectra and we extract $\Ep$ and $\Gamma$ using Gaussian fits. 
For sufficiently weak interactions, we observe a linear dependence of $\Ep$ on~$a$ [Fig.~\ref{fig8}(b), top]. 
We calibrate the effective volume of our two boxes so that a linear fit to the data with shifts $<1$\,kHz recovers the mean-field result $E_{\rm p} = gn$, with \mbox{$g = 2 \pi \hbar^2 a/m_r$}; note that we also independently correct for small changes in the effective box volume based on the chemical potential of the bath.

The half-width $\Gamma$~\cite{residualtrf} is also roughly proportional to $gn$ [see Fig.~\ref{fig8}(b), bottom]. This unexpected dependence is heuristically captured by $\Gamma=0.64|g|n$ (solid line).

\vspace{-1em}
\setcounter{equation}{0} 
\renewcommand\theequation{D\arabic{equation}} 
\section*{Appendix~D:~Mean-field simulations}
\vspace{-1em}
\begin{figure}[b!]
\centerline{\includegraphics[width=1\columnwidth]{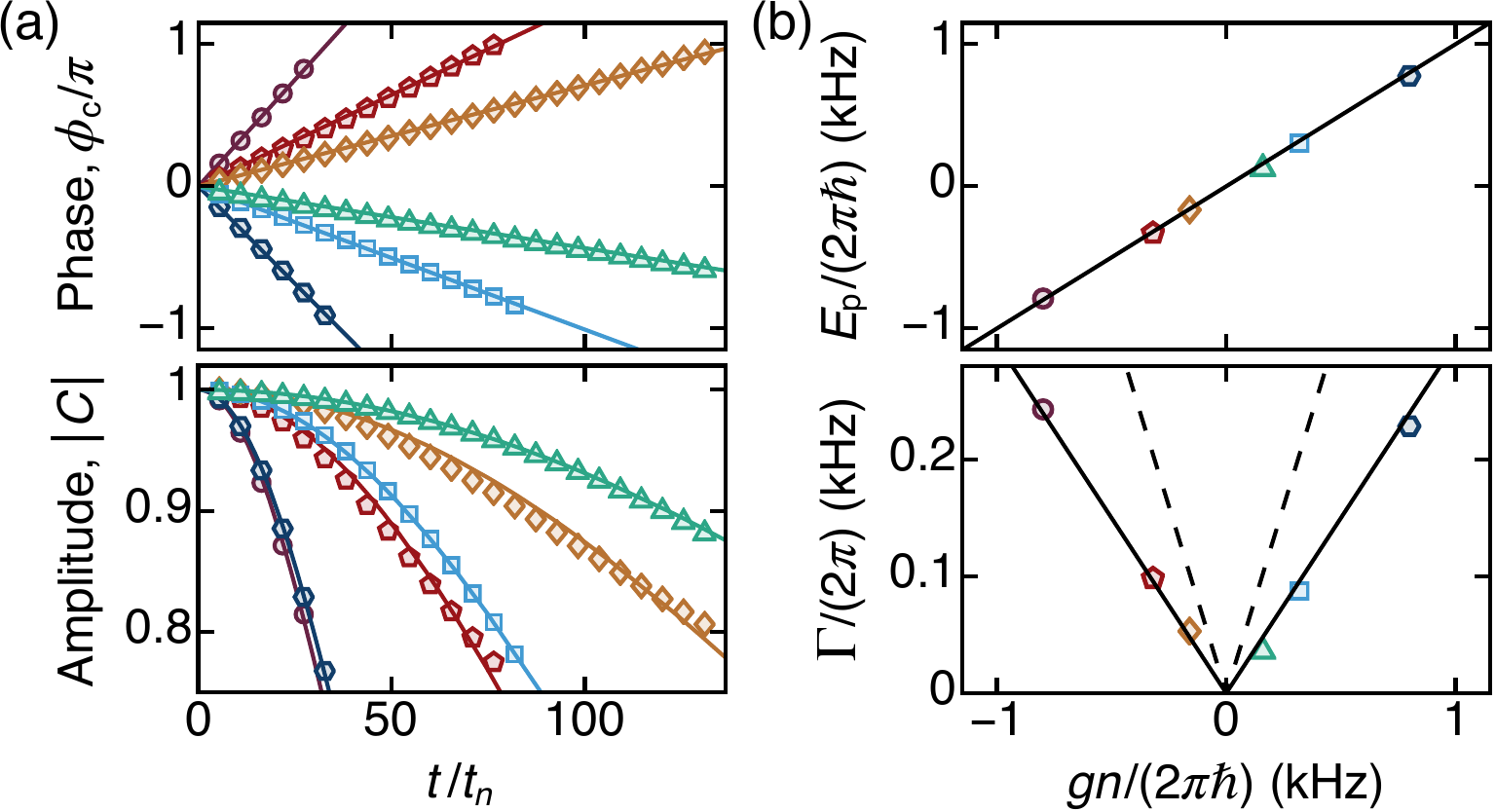}}
\caption{
Mean-field simulations of Ramsey interferometry. (a)~Phase and amplitude of $C(t)$ for varying $a$ [indicated by the symbols, see (b)].
The solid lines show fits to $\phi_{\rm c}$ and $|C|$.
(b)~Extracted $\Ep$ (top) and rate of decoherence $\Gamma$ (bottom) versus $gn$ (see text). The solid lines show $\Ep=gn$ (top) and $\Gamma=0.3|g|n$ (bottom). The dashed line shows $\Gamma=0.64|g|n$ from experiments.
}
\label{fig9}
\end{figure}

We simulate our system on the mean-field level using the two-component Gross--Pitaevskii equation
\be
\begin{split}
i\hbar\frac{\partial \psi_\uparrow}{\partial t}= -\frac{\hbar^2}{2m}\nabla^2 \psi_\uparrow+g_{\uparrow\uparrow}|\psi_\uparrow|^2\psi_\uparrow
+g_{\uparrow\downarrow}|\psi_\downarrow|^2\psi_\uparrow,\\
i\hbar\frac{\partial \psi_\downarrow}{\partial t}= -\frac{\hbar^2}{2m}\nabla^2 \psi_\downarrow+
g_{\downarrow\downarrow}|\psi_\downarrow|^2\psi_\downarrow
+g_{\uparrow\downarrow}|\psi_\uparrow|^2\psi_\downarrow,
\end{split}
\label{GPE}
\ee
with $g_{\uparrow\uparrow}= 4 \pi \hbar^2 \ab/m$, $g_{\downarrow\downarrow}= 4 \pi \hbar^2 \ai/m$, and $g_{\uparrow\downarrow}=g= 2 \pi \hbar^2 a/m_r$.
We use a pseudo-spectral method with fourth-order Runge--Kutta time evolution to solve Eqs.~(\ref{GPE}), performed on a $64\times64\times128$ grid with dimensions $40\times40\times80\,\upmu\textrm{m}^3$ and a $4\,\upmu\textrm{s}$ time-step.
Our simulations start from the ground-state wavefunction of $N$ particles in $\ket{\uparrow}$ state, $\psi_0$, obtained using imaginary-time evolution. To obtain $C(t)$, we initiate the real-time dynamics with $\psi_{\uparrow}=\sqrt{1-f}\psi_0$ and $\psi_{\downarrow}=\sqrt{f}\psi_0$, where $f$ is the impurity fraction, wait for an evolution time $t$, and calculate $C(t)=\braket{\psi_{\downarrow}(t)|\psi_{\uparrow}(t)}/\braket{\psi_{\downarrow}(0)|\psi_{\uparrow}(0)}$.
Note that for all experimental combinations of $a$, $\ab$, $\ai$, the two-component mixture is mean-field unstable; at long times, we observe signatures of wavefunction collapse and associated numerical instability, but initially, the dynamics are well-defined. 

In Fig.~\ref{fig9}(a), we show the evolution of $C(t)$ for varying $a$, with $\ab=18a_0$, $\ai=-64a_0$, $N=2.7\times10^5$, $f=0.05$, and a box potential with $R=15\,\upmu$m, $L=50\,\upmu$m, and depth of $50$\,nK.
We estimate $\Ep$ and $\Gamma$ using linear and Gaussian fits to $\phi_{\mathrm{c}}$ and $|C|$, respectively.
The extracted $\Ep$ (corrected for the mean-field shift of the bath, $\hbar \delta_{\rm mf}$) is consistent with $\Ep=gn$, where $n=\int|\psi_\uparrow|^4\mathrm{d^3}\mathbf{r}/N$ is the average bath density. We find $\Gamma\approx 0.3 |g|n$, which captures the experimentally observed $\Gamma\propto|g|n$ scaling [see Fig.~\ref{fig8}(b)], but not its absolute value.
We attribute this decoherence to dynamics arising from the sudden quench of an effective potential felt by the impurities, given by $gn(\mathbf{r})=g|\psi_\uparrow(\mathbf{r})|^2$, which varies over the healing length $\xi=1/\sqrt{8\pi n \ab}$ near the box edges (in Fig.~\ref{fig9}, $\xi\approx2\,\upmu$m).
We note that such dynamics distributes the impurities over various momenta, and this could lead to further broadening by mechanisms not captured in the GPE (see~e.g.~\cite{Seetharam:2021}).

We have also checked that we obtain essentially the same results with $\ai=0$ and $\ai=+64a_0$, indicating that for small transfer fractions the impurity-impurity interactions do not play a significant role.

\vspace{-1em}
\setcounter{equation}{0} 
\renewcommand\theequation{E\arabic{equation}} 
\section*{Appendix~E:~Early-time $C(t)$ for $1/(k_n a)\leq0$}
\vspace{-1em}
{
In Fig.~\ref{fig10} we show the early-time evolution of $C(t)$, for $1/(k_n a)\leq0$ data from Fig.~\ref{fig2}~(c), plotted versus $(t/t_n)^{3/2}$.
The dashed lines show the $a$-independent early-time prediction (valid for \mbox{$t/t_n\to0$})\cite{Parish:2016}, see also \cite{Skou:2021}:
\be
\begin{split}
  |C(t)|&=1-\frac{16}{9\pi^{3/2}}\left(\frac{t}{t_n}\right)^{3/2},\\
  \phi_{\rm c}(t)&= \frac{16}{9\pi^{3/2}}\left(\frac{t}{t_n}\right)^{3/2}\,,
\end{split}
\label{eq:earlytimes}
\ee
which also coincides with the $t/t_n\to 0$ behavior of the analytical prediction of Ref.~\cite{Drescher:2021} in Eq.~(\ref{eq:ideal}).

\begin{figure}[h!]
\centerline{\includegraphics[width=1\columnwidth]{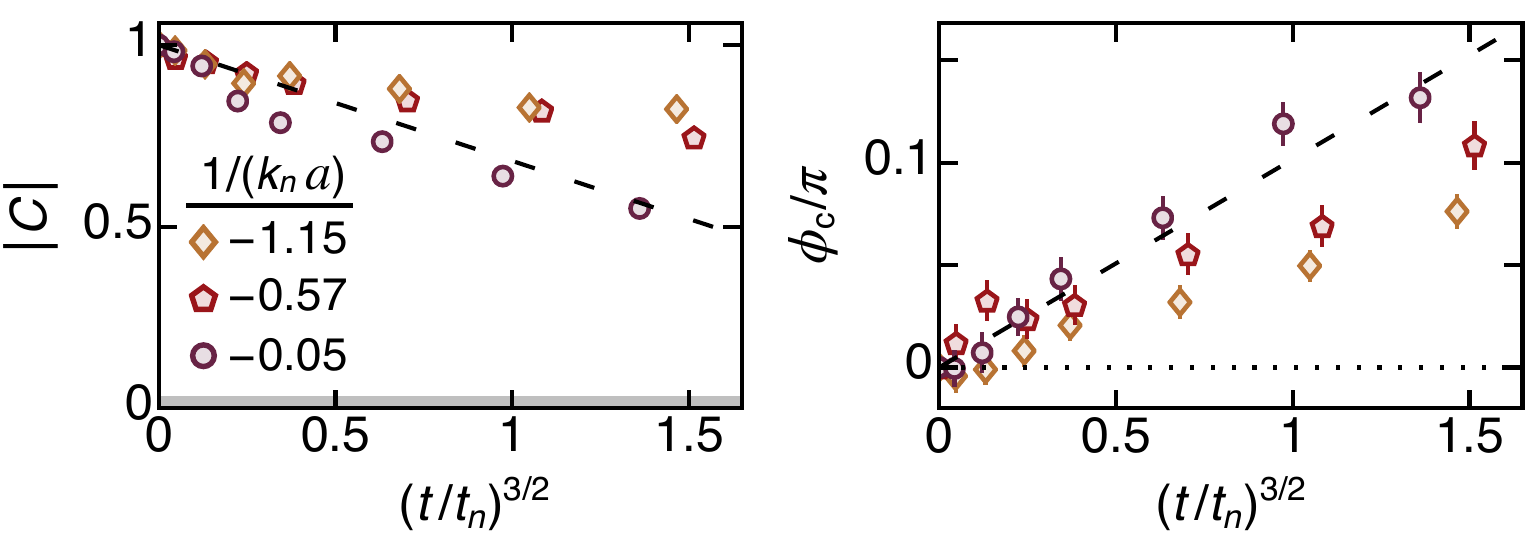}}
\caption{
Early-time $C$ dynamics for attractive interactions [zoom-in on the data from Fig.~\ref{fig2}(c) plotted versus $(t/t_n)^{3/2}$]. 
The dashed lines show the $a$-independent early-time ($t/t_n\ll1$)  predictions from Eqs.~(\ref{eq:earlytimes}). The error bars show fitting errors.
}
\label{fig10}
\end{figure}

\setcounter{equation}{0} 
\renewcommand\theequation{F\arabic{equation}} 
\section*{Appendix~F:~Atom loss at unitarity}

To measure the atom loss at unitarity we perform a short rf pulse at weak interactions and then quench the field to unitarity for a variable hold time (this protocol is equivalent to the Ramsey protocol [Fig.~\ref{fig2}(a)] without the second pulse).
We do not resolve impurities and bath atoms separately (our low-field absorption imaging is not state selective), but instead measure the total fractional atom loss $\Delta N/N$. We estimate the decoherence due to this loss using:
\be
|C(t)|= \sqrt{1-\frac{\Delta N(t)}{\Delta N_{\rm max}}}\,,
\label{eq:Closs}
\ee  
where $\Delta N_{\rm max}$ is the number of atoms lost at long times (when all impurities have been lost), which assumes that the fractional loss rate is much higher for impurities than bath atoms (satisfied for small impurity fractions).

In Fig.~\ref{fig11} we show the $|C|$ measurements at unitarity from Fig.~\ref{fig3} and the expected decoherence due to atom loss alone (solid symbols), calculated using Eq.~(\ref{eq:Closs}). For $t/t_n>1$, both are empirically captured well by decaying exponentials (solid lines), with inverse lifetimes (seen in the slopes) that differ by a factor of $4$, which shows that loss only plays a small role.

\setcounter{equation}{0} 
\renewcommand\theequation{G\arabic{equation}}
\section*{Appendix~G:~
Further data on the universality of the Bose polaron spectrum
}

\begin{figure}[t!]
\centerline{\includegraphics[width=1\columnwidth]{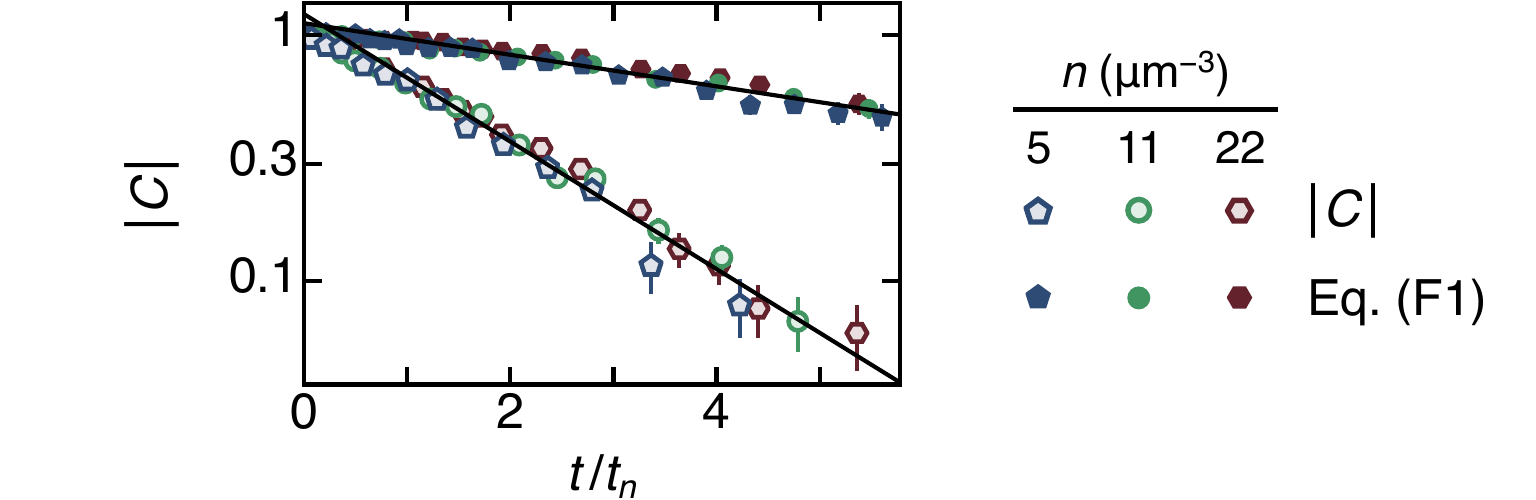}}
\caption{
Assessing decoherence due to atom loss. The open symbols show data from Fig.~\ref{fig3}, while the solid ones show an estimate of the decay of $|C|$ due to atom loss [calculated using Eq.~(\ref{eq:Closs})].
The solid lines show decaying exponential fits to $t/t_n>1$. The error bars show fitting errors.
}
\label{fig11}
\end{figure}

\begin{figure}[b!]
\centerline{\includegraphics[width=1\columnwidth]{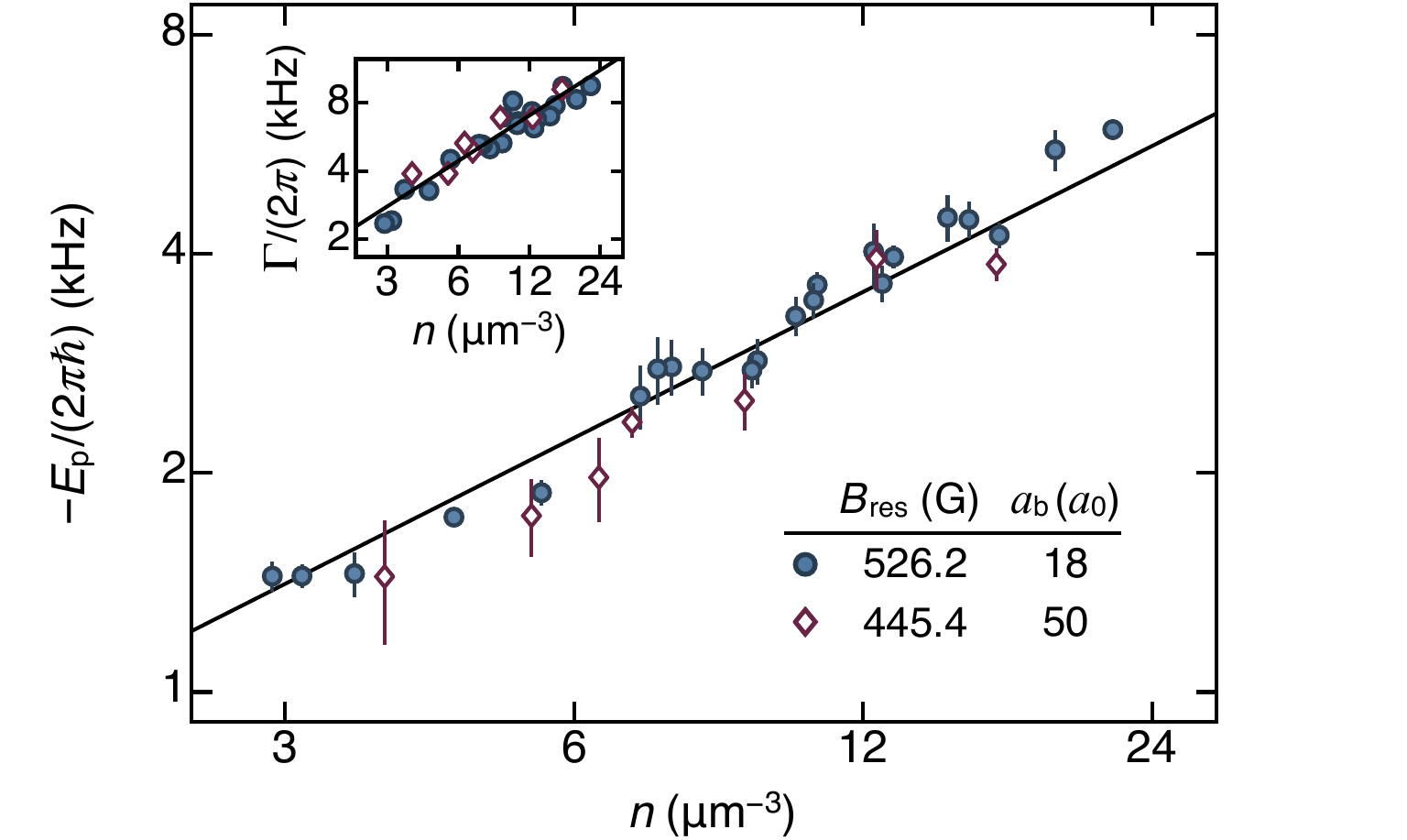}}
\caption{
Scaling laws at unitarity.
Extracted mode $\Ep$ (main panel) and width $\Gamma$ (inset) of $I(\delta)$ as a function of density $n$ and for our two $\ab$ (legend). 
The solid lines show $\Ep= -0.34 E_n$ and $\hbar\Gamma=0.68 E_n$. The error bars reflect fitting errors.
}
\label{fig12}
\end{figure}
 
\vspace{-1em}
\subsection*{Scaling of $\Ep$ and $\Gamma$ at unitarity}
\vspace{-1em}
Complementing Fig.~\ref{fig3},  in Fig.~\ref{fig12} we explicitly verify the universal $\propto n^{2/3}$ scaling at unitarity using spectroscopic measurements for both Feshbach resonances.
We show the data from Fig.~\ref{fig4} near unitarity [$-0.2\lesssim1/(k_n a)\lesssim0.2$], plotting $\Ep$ versus $n$ for the two $\ab$ values on log-log scale. We observe no discernible dependence on $\ab$ (within errors), and a power-law fit $\Ep\propto n^\gamma$, gives $\gamma=0.7(1)$.
Similarly, we also find that $\Gamma\propto n^\gamma$ with $\gamma=0.7(1)$ (see inset). These results are consistent with $\gamma=2/3$, expected from the density-set $E_n$ scaling. Assuming $\gamma=2/3$, we find $\Ep=0.34(1) E_n$ and $\Gamma=0.68(1) E_n$ [solid lines in Fig.~\ref{fig12}].
These results further imply that in our parameter range the spectral function can only depend weakly on $\ab$ and other lengthscales. 

\vspace{-1em}
\subsection*{Comparison of dynamics for different $n$}
\vspace{-1em}
In Fig.~\ref{fig13}, we further show that $C(t)$ for different $n$ but the same $1/(k_n a)$ (at two characteristic values) are identical.

\begin{figure}[h!]
\centerline{\includegraphics[width=1\columnwidth]{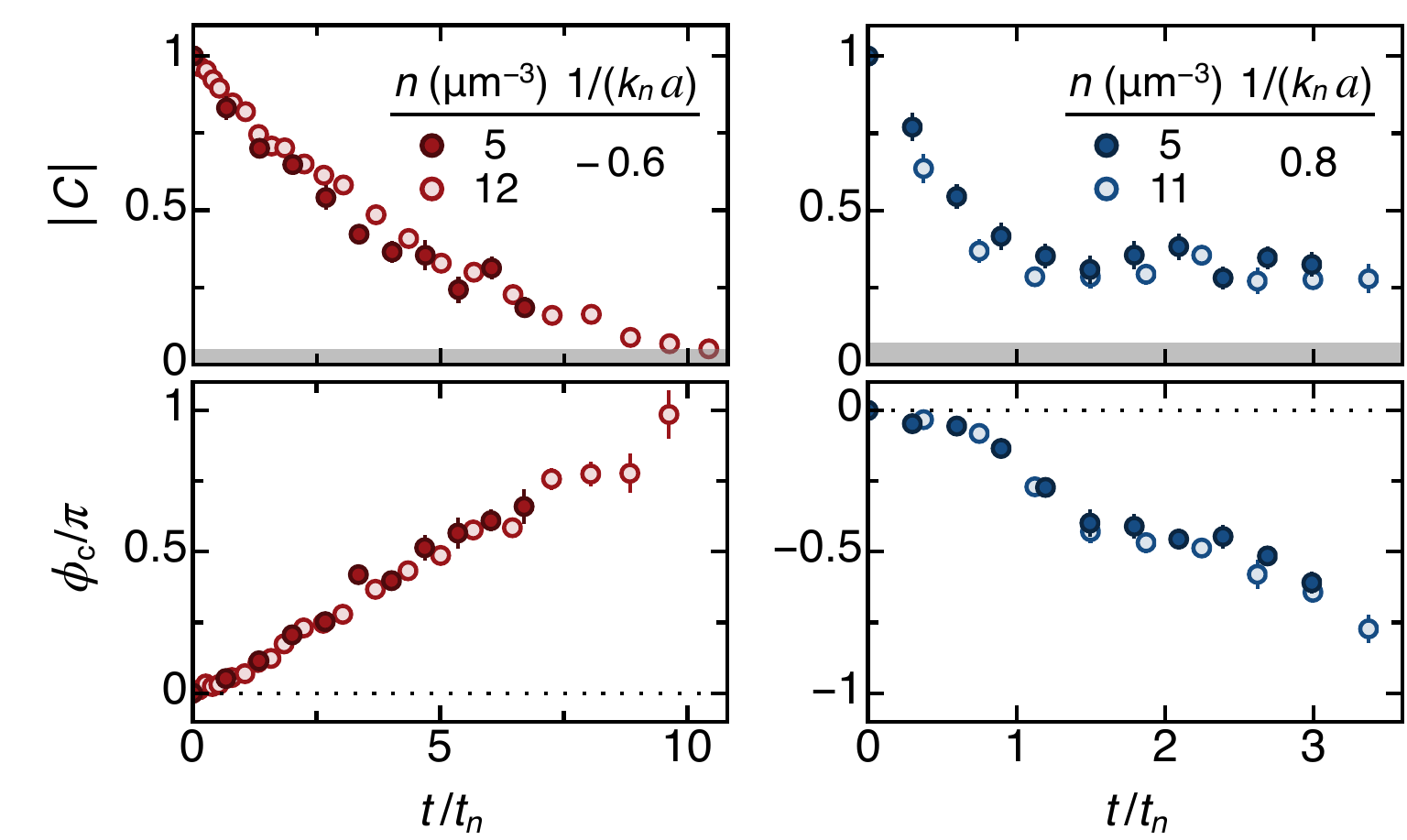}}
\caption{
Comparison of $|C|$ and $\phi_{\rm c}$ for different $n$ (legend) but the same $1/(k_n a)$ (left: $-0.6$, right: $+0.8$); for analogous plots at $1/(k_n a)\approx 0$ see Fig.~3. The error bars show fitting errors.
}
\label{fig13}
\end{figure}

\vspace{-2em}
\subsection*{Comparison of $\Ep$ and $\Gamma$ for spectroscopy and interferometry}
\vspace{-1em}

In Fig.~\ref{fig14} we compare $\Ep$ and $\Gamma$ values for the polaron branches extracted from the Fourier transforms of $C(t)$ data, to those obtained from spectroscopy (Fig.~\ref{fig4}), complementing the comparison of full spectra at three characteristic $1/(k_n a)$ in Fig.~\ref{fig2}(d). 
We observe good agreement, with only small systematic differences discernible at unitarity and for strong repulsive interactions, and note that the differences in $\Ep$ are a small fraction of $\hbar\Gamma$.
While interferometry is intrinsically free of Fourier broadening~\cite{residualtrf}, spectroscopy is better for resolving distinct peaks in the frequency domain, and so here we do not attempt to extract $\Ep$ and $\Gamma$ for the attractive branch at $1/(k_na)\gtrsim0.5$ using interferometry~[cf. Fig.~\ref{fig2}(d) and Fig.~\ref{fig5}].

\begin{figure}[h!]
\centerline{\includegraphics[width=1\columnwidth]{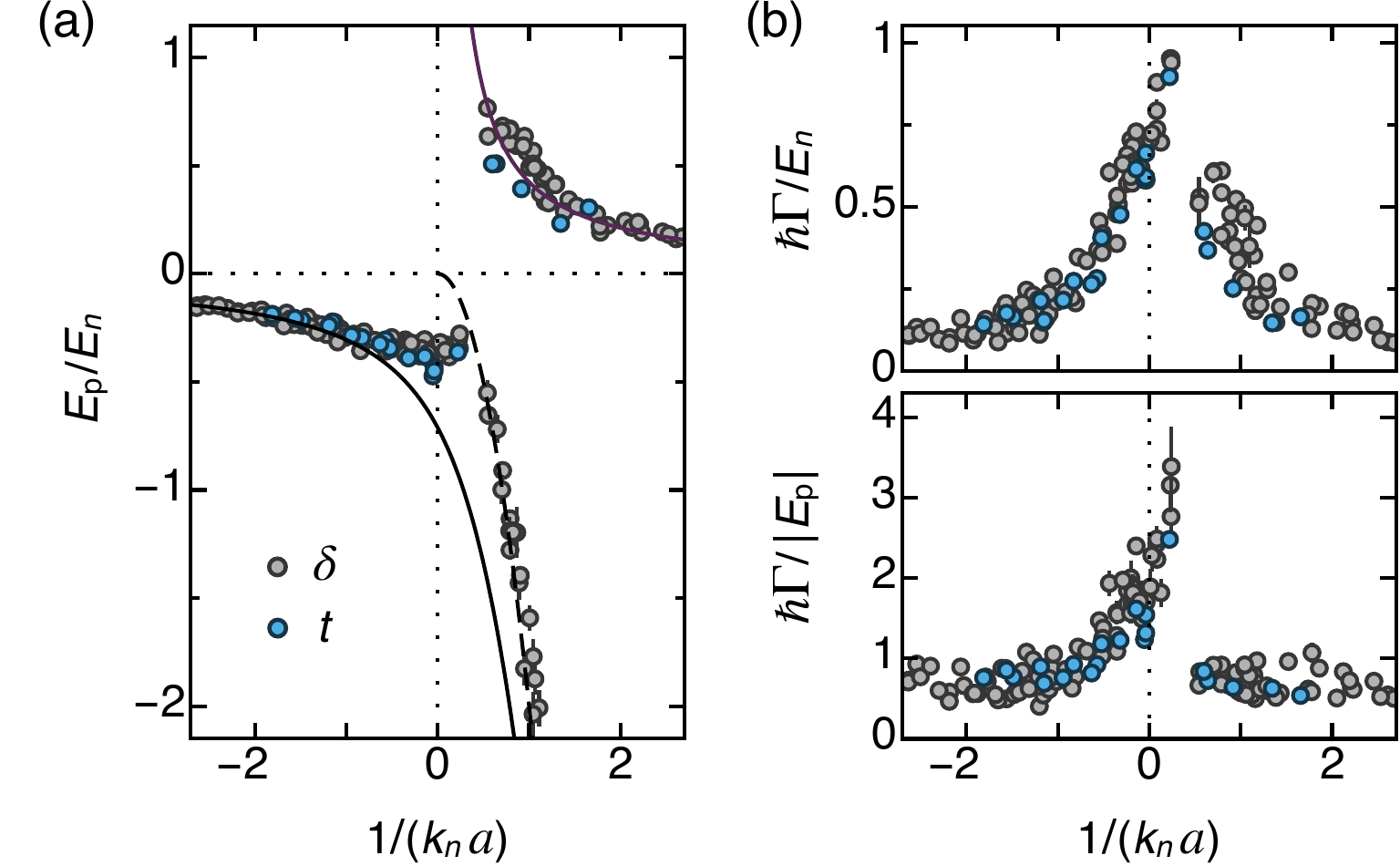}}
\caption{
Comparison of extracted $\Ep$ and $\Gamma$ from spectroscopy and interferometry.
We reproduce panels from Fig.~\ref{fig4}, showing the spectroscopy data for $\Bres=526.2\,$G (gray points). 
The blue points show $\Ep$ and $\Gamma$ extracted from the Fourier transforms of $C(t)$ [see \emph{e.g.} Fig.~\ref{fig2}(d)]. The error bars reflect fitting errors.
}
\label{fig14}
\end{figure}

\vspace{-1em}
\subsection*{Comparison with harmonic-trap experiments}
\vspace{-1em}
In Fig.~\ref{fig15} we compare our $\Ep$ values to measurements from Refs.~\cite{Hu:2016,Jorgensen:2016,Morgen:2025}.
For the JILA experiment~\cite{Hu:2016}, we only include points where the peak position could be unambiguously fitted. For the Aarhus experiments~\cite{Jorgensen:2016, Morgen:2025}, we include data for $1/(k_n a)\lesssim 0$ from Ref.~\cite{Jorgensen:2016} and for $1/(k_n a)> 0$ from~\cite{Morgen:2025}; note that Ref.~\cite{Jorgensen:2016} also includes measurements for $1/(k_n a)> 0$, but they are superseded by the  measurements in Ref.~\cite{Morgen:2025}.

When plotted in dimensionless form and versus $1/(k_n a)$, the data show no systematic differences, despite significant variations in the parameters; specifically, $\ab=100a_0$ in~\cite{Hu:2016} and $9a_0$ in~\cite{Jorgensen:2016,Morgen:2025}, and typical average $E_n/(2\pi \hbar)\approx 40\,$kHz in~\cite{Hu:2016,Morgen:2025} and $\approx70\,$kHz in~\cite{Jorgensen:2016}.
The different mass ratio in \cite{Hu:2016} ($^{40}$K impurities in a $^{87}$Rb BEC) enters only via $m_r$ in the definition of $E_n$.

\begin{figure}[t!]
\centerline{\includegraphics[width=1\columnwidth]{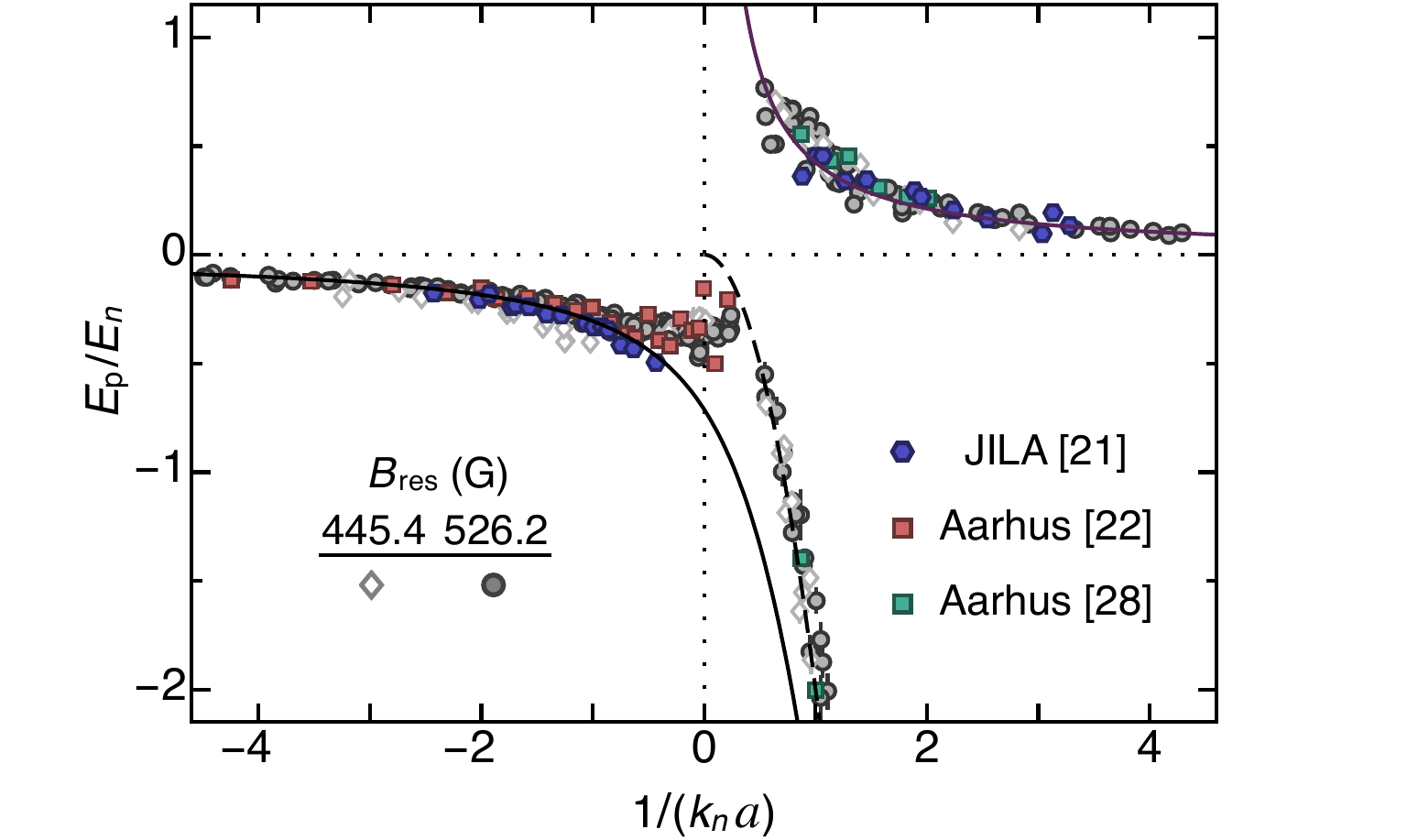}}
\caption{
Comparison of extracted $\Ep/E_n$ versus $1/(k_n a)$
between our measurements [gray,~from Fig.~\ref{fig4}(a)] and previous harmonic-trap injection experiments (omitting error bars for clarity).
}
\label{fig15}
\end{figure}

\setcounter{equation}{0} 
\renewcommand\theequation{H\arabic{equation}} 
\section*{Appendix~H:~Beating of the two branches for $1/(k_n a)= 0.6$}

In Fig.~\ref{fig16}, we show the raw $C(t)$ data for $\tilde{I}_{\rm \pm}$ in Fig.~\ref{fig5}(c). For the quench from $a<0$ we observe clear beating between the two states, with characteristic $\pi$-jumps in phase when $|C(t)|$ nears zero (see also \cite{Cetina:2016}). 

\begin{figure}[h!]
\centerline{\includegraphics[width=1\columnwidth]{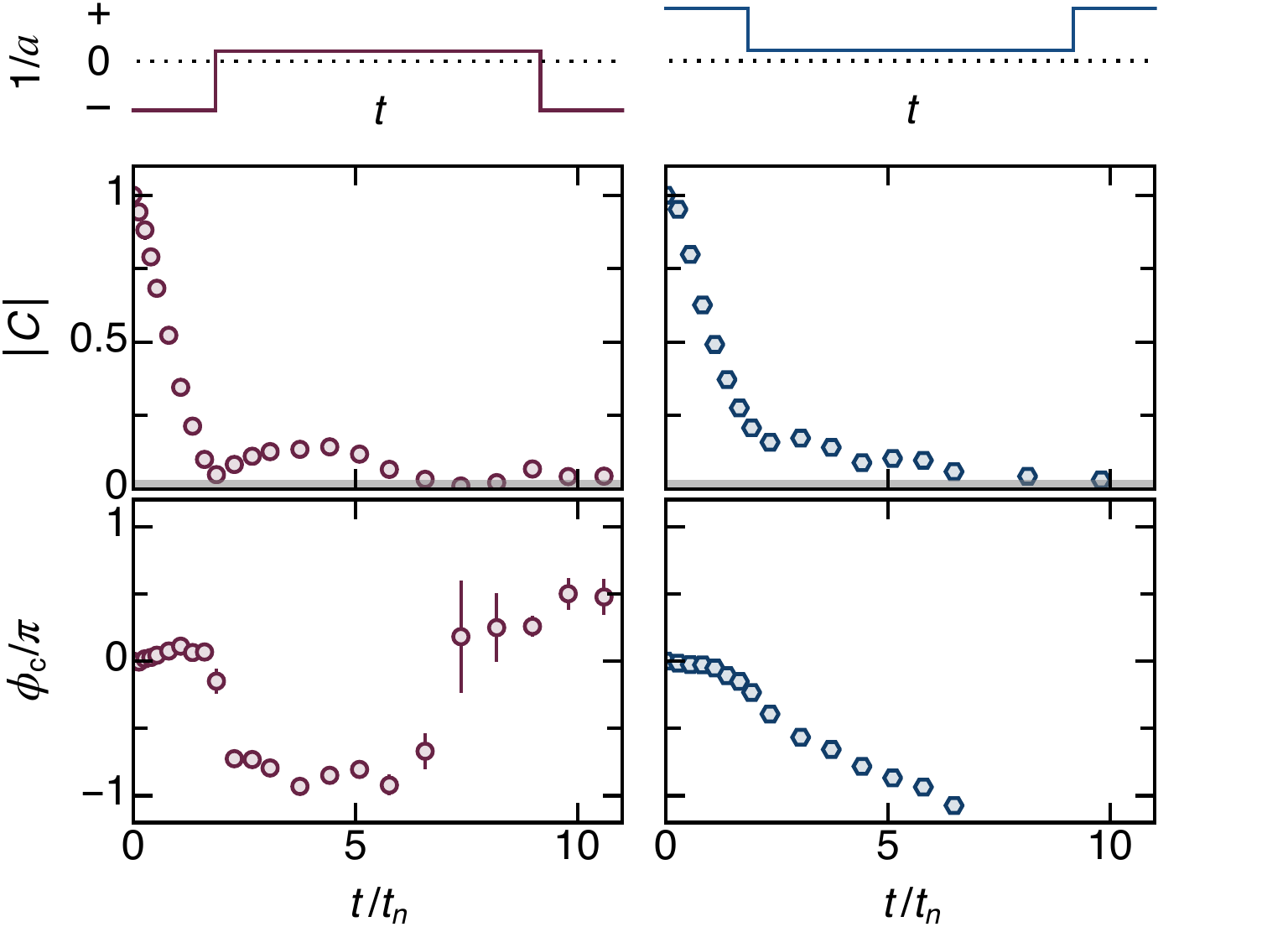}}
\caption{
Dynamics of $C(t)$ for the two measurements from Fig.~\ref{fig5}(c), using quenches from either $a<0$ (left) or $a>0$ (right). The error bars show fitting errors.
}
\label{fig16}
\end{figure}


\begin{thebibliography}{81}
\bibitem [{win()}]{windingfn}%
  \BibitemOpen
  \href@noop {} {}\bibinfo {note} {Note that we account for the additional phase
  winding during the evolution time due to $\omega_0$ changing with
  $B$.}\BibitemShut {Stop}%
\bibitem [{res()}]{residualtrf}%
  \BibitemOpen
  \href@noop {} {}\bibinfo {note} {We correct $\Gamma$ for residual
  finite-$t_{\rm rf}$ broadening, see Appendix~B for details.}\BibitemShut
  {Stop}%
\bibitem [{ext()}]{extralengthscales}%
  \BibitemOpen
  \href@noop {} {}\bibinfo {note} {Note that our experiments probe the case of
  broad Feshbach resonances and equal impurity and bath-atom mass. For a narrow
  resonance, the interaction range provides an additional relevant lengthscale,
  and for light impurities the Efimov effect is expected to play a significant
  role.}\BibitemShut {Stop}%
\bibitem [{Epc()}]{Epcomment}%
  \BibitemOpen
  \href@noop {} {}\bibinfo {note} {For the polaron branches, when
  $\hbar\Gamma/|\Ep|$ is essentially constant, we interpret $\Ep$ as the
  polaron ground-state energy. Near unitarity, significant weight is
  transferred to excited states and $\Ep$ reflects their weighted average.
  Further elucidating and possibly eliminating the sources of the observed
  symmetric broadening remains an important task for the future.}\BibitemShut
  {Stop}%
\bibitem [{str()}]{strongrf}%
  \BibitemOpen
  \href@noop {} {}\bibinfo {note} {Note that here we resort to stronger rf
  pulses to improve the signal of the molecular branch, potentially
  compromising the linear response regime for the repulsive
  polaron.}\BibitemShut {Stop}%
\end{thebibliography}
\end{document}